\documentclass[prd,showpacs,superscriptaddress,nofootinbib,floatfix,12pt]{revtex4-1} 
\usepackage[utf8]{inputenc}
\usepackage{amsmath,amssymb}
\usepackage{slashed}
\usepackage{graphicx}
\usepackage{subfigure}
\usepackage[colorlinks]{hyperref}
\usepackage[usenames,dvipsnames]{color}

\usepackage{mathrsfs}
\usepackage{color}
\usepackage{bbm}               
\usepackage{xspace}				
\usepackage{epstopdf}
\usepackage{pgffor}							

\newcommand*{\intMC}{\frac{1}{\text{flux}}\sum_{\text{all}}\int}

\newcommand*{\pairMk}{\frac{d^3k}{(2\pi)^32\mu}}

\newcommand*{\all}{\text{all}}

\newcommand*{\Tr}{\text{Tr}}
\newcommand*{\geff}{g_\text{eff}}
\usepackage{newtxtext, newtxmath}
\usepackage{braket,bm}
\usepackage{multirow}
\usepackage{enumitem}

\newcommand{\itp}{\affiliation{CAS Key Laboratory of Theoretical Physics, Institute of Theoretical Physics,\\ Chinese Academy of Sciences, Beijing 100190, China}}

\newcommand{\ucas}{\affiliation{School of Physical Sciences, University of Chinese Academy of Sciences, Beijing 100049, China}}

\newcommand{\uestc}{\affiliation{School of Physics, University of Electronic Science and
Technology of China,\\ Chengdu 610054, China}} 

\begin{document}

\title{Semi-inclusive electroproduction of hidden-charm and double-charm hadronic molecules}

\author{Pan-Pan Shi}
\email{shipanpan@itp.ac.cn}
\author{Feng-Kun Guo} \email{fkguo@itp.ac.cn}
\itp \ucas

\author{Zhi Yang}\email{zhiyang@uestc.edu.cn} \uestc

\begin{abstract}
    The semi-inclusive electroproduction of exotic hadrons, including the $T_{cc}$, $P_{cs}$, and hidden-charm baryon-antibaryon states, is explored under the assumption that they are $S$-wave hadronic molecules of a pair of charmed hadrons. We employ the Monte Carlo event generator Pythia to produce the hadron pairs and then bind them together to form hadronic molecules. With the use of such a production mechanism, the semi-inclusive electroproduction rates are estimated at the order-of-magnitude level. Our results indicate that a larger number of $P_{cs}$ states and $\Lambda_c\bar{\Lambda}_c$ molecules can be produced at the proposed electron-ion colliders in China (EicC) and in the US (EIC). The results also suggest that the $T_{cc}$ states and other hidden-charm baryon-antibaryon states can be searched for at EIC. Besides, the potential 24-GeV upgrade of the Continuous Beam Accelerator Facility at the Thomas Jefferson National Accelerator Facility can play an important role in the search for the hidden-charm tetraquark and pentaquark states due to its high luminosity. 
\end{abstract}

\date{\today}
\maketitle

\section{Introduction}

In the last two decades, a number of hadron structures have been observed in various high-energy experiments, such as BESIII~\cite{BESIII:2020nme}, LHCb~\cite{Cerri:2018ypt}, Belle~\cite{Belle-II:2018jsg}. Many of these structures possess properties incompatible with the predictions of the traditional quark model for quark-antiquark mesons and three-quark baryons, and therefore are excellent candidates for the so-called exotic hadrons. These structures have been studied extensively with different models, but debates about their nature still exist (for recent reviews, see Refs.~\cite{Chen:2016qju,Hosaka:2016pey,Lebed:2016hpi,Esposito:2016noz,Guo:2017jvc,Olsen:2017bmm,Liu:2019zoy,Brambilla:2019esw,Guo:2019twa,Chen:2022asf}).

So far, most of the exotic states have been observed at hadron-hadron and electron-positron collisions. To understand the nature of the exotic states, other production processes have been proposed to search for these states. Notably, photoproduction or leptoproduction processes have the advantage that the possible resonant signals are free of triangle singularities (for a review, see Ref.~\cite{Guo:2019twa}). Pioneering work in the search for exotic hadrons in photoproduction processes has been done by the COMPASS and GlueX collaborations. The COMPASS Collaboration searched for $Z_c(3900)^{\pm}$ in the  $J/\psi\pi^{\pm}$ invariant-mass distribution, but there was no signal~\cite{COMPASS:2014mhq}. Later, the COMPASS Collaboration observed a peak around 3.86~GeV in the $J/\psi\pi^+\pi^-$ invariant-mass distribution with a $4.1\sigma$ statistical significance, the $\tilde{X}(3872)$, in muoproduction process~\cite{COMPASS:2017wql}, and the measured $\pi^+\pi^-$ invariant-mass distribution indicates a negative $C$ parity for this structure. The GlueX Collaboration found no evidence of hidden-charm pentaquark $P_c$ states in near-threshold $J/\psi$ exclusive photoproduction off the proton in Hall D at Thomas Jefferson National Accelerator Facility (JLab) and set model-dependent upper limits (using a vector-meson-dominance model) for the branching fractions $\text{Br}[P_c^+\to J/\psi p]$~\cite{GlueX:2019mkq}.

The proposed electron-ion collider in China (EicC)~\cite{Anderle:2021wcy}, the Electron-Ion Collider in the US (EIC)~\cite{AbdulKhalek:2021gbh}, and a potential 24-GeV upgrade of the Continuous Beam Accelerator Facility (CEBAF) at JLab~\cite{Arrington:2021alx} provide new opportunities to search for exotic hadron structures and study the nature of such states. The center-of-mass (c.m.) energies at EicC and EIC are much higher than the thresholds of charmed hadron pairs, and thus they permit semi-inclusive electroproduction of hidden-charm (and even double-charm) exotic states. 
For CEBAF (24~GeV), the c.m. energy of the electron-proton system with a 24~GeV electron beam hitting the proton target is about 6.7~GeV and thus much lower than that at EicC and EIC; yet, its luminosity is much higher.
The energy configurations and luminosity of these facilities are listed in Table~\ref{Tab:Lum_EicC_EIC}.

Many works have estimated the cross sections for the exclusive photoproduction of the hidden-charm pentaquark states~\cite{Huang:2013mua,Wang:2015jsa,Huang:2016tcr,HillerBlin:2016odx,Karliner:2015voa,Kubarovsky:2015aaa,Winney:2019edt,Paryev:2018fyv,Wang:2019krd,Goncalves:2019vvo,Wu:2019adv,Xie:2020niw,Yang:2020eye} and hidden-charm tetraquark states~\cite{Liu:2008qx,Galata:2011bi,Lin:2013mka,Lin:2013ppa,Wang:2015lwa,Albaladejo:2020tzt} with the vector-meson dominance model, which assumes that the photon emitted from the electron converts into a $J/\psi$ which then interacts with the proton to produce the hidden-charm states. 
Yet, since the production of a heavy quarkonium, such as $J/\psi$, in high-energy reactions requires the heavy quark and antiquark pair to be confined in a small phase space, its production rate is much lower than that of a pair of open-charm hadrons.
In view of this, the open-charm hadron pair might be a crucial component in producing exotic hadrons with hidden-charm.
In particular, in Ref.~\cite{Du:2020bqj} it was pointed out that the total cross section for the near-threshold photoproduction of $J/\psi$ in $\gamma J/\psi \to J/\psi p$ measured by GlueX is consistent with assuming that the reaction is dominated by the $\Lambda_c \bar{D}^{(*)}$ channels through $\gamma J/\psi \to \Lambda_c \bar{D}^{(*)} \to J/\psi p$. 
Furthermore, the results for the heavy vector quarkonia, calculated by solving the Dyson-Schwinger equation, are incompatible with the assumption of the vector-meson-dominance model in the electroproduction process~\cite{Xu:2021mju}. Therefore, it is necessary to explore the photoproduction of exotic states with other methods.

Based on the hadronic molecular picture, the semi-inclusive leptoproduction rates of the hidden-charm hadrons at COMPASS, EicC, and EIC were estimated in Ref.~\cite{Yang:2021jof}. The considered production mechanism is such that the charmed hadron pairs are semi-inclusively generated with the use of a Monte Carlo (MC) event generator and then bound together to form hadronic molecules through the final-state interaction (FSI)~\cite{Artoisenet:2010uu,Guo:2014ppa,Guo:2014sca}. 
This mechanism has been employed to estimate the cross sections at the order-of-magnitude level for the production of the $X(3872)$~\cite{Bignamini:2009sk,Artoisenet:2009wk,Albaladejo:2017blx}, the spin partner and bottom analogues of $X(3872)$~\cite{Guo:2014sca}, the charm-strange hadronic molecules~\cite{Guo:2014ppa}, the charged charmonium-like and bottomonium-like states~\cite{Guo:2013ufa}, the $P_c$ states~\cite{Ling:2021sld}, and dionium ( $D^+D^-$ hadronic atom)~\cite{Shi:2021hzm}. The so-obtained cross sections for the production of $X(3872)$~\cite{Guo:2017jvc} at hadron colliders are in line with the measurements at CDF~\cite{Bauer:2004bc} and CMS~\cite{CMS:2013fpt}.

In this work, we employ the same mechanism to estimate the cross sections for the semi-inclusive electroproduction of the double-charm $T_{cc}$ states, the hidden-charm $P_{cs}$ states, and hidden-charm baryon-antibaryon states at EicC and EIC, which have not been given before, and the semi-inclusive electroproduction rates of $X(3872)$, $Z_c(3900)^{0(+)}$, the $P_c$ states, and $P_{cs}(4459)$ at JLab for the first time.\footnote{The cross sections for the production of the $P_c$ states are significantly small due to the low c.m. energy (up to about 4.8 GeV) at the running JLab experiments. In this work, we only discuss the possible upgrade of CEBAF with the electron energy of 24~GeV.} Among the exotic states discussed in this work, $P_c(4312)$, $P_c(4440)$, $P_c(4457)$, $P_{cs}(4459)$, and $T_{cc}^+$ were observed by the LHCb Collaboration~\cite{LHCb:2019kea,LHCb:2020jpq,LHCb:2021auc,LHCb:2021vvq}, $X(3872)$ was first reported by the Belle Collaboration~\cite{Belle:2003nnu}, $Z_c(3900)^{+}$ was discovered by the BESIII and Belle Collaborations~\cite{BESIII:2013ris,Belle:2013yex}, and $Z_c(3900)^0$ was observed by the BESIII Collaboration~\cite{BESIII:2015cld}. The rest of the states are predicted in the hadronic molecular picture; see Refs.~\cite{Dong:2021juy,Dong:2021bvy} for a survey of the spectrum of hidden-charm and double-charm hadronic molecules.

This paper is organized as follows. In Sec.~\ref{Sec:general_machanism}, we introduce the  semi-inclusive electroproduction mechanism of hadronic molecules. Our numerical results and the production rates for the exotic states at EicC, EIC, and the 24~GeV upgrade of CEBAF are presented in Sec.~\ref{Sec:result}. We briefly summarize our work in Sec.~\ref{Sec:summary}. The Appendix contains the derivation of the production cross section of the spin-$3/2$ molecule, which is composed of a baryon and a vector meson.

\begin{table}[tb]
\caption{\label{Tab:Lum_EicC_EIC} Energy configurations and luminosities for EicC, EIC, and the proposed 24~GeV upgrade of CEBAF. The integrated luminosities correspond to about 1 year of operation. }
\renewcommand{\arraystretch}{1.2}
\begin{tabular*}{\columnwidth}{@{\extracolsep\fill}lccc}
\hline\hline                                           
           & EicC~\cite{Anderle:2021wcy}   & EIC~\cite{AbdulKhalek:2021gbh}    & CEBAF (24 GeV)~\cite{Arrington:2021alx} \\[3pt]
\hline
    $e$ [GeV]      & $3.5$ & $18$  & $24$ \\[3pt]
    $p$ [GeV]      & $20$ & $275$  & $0$ \\[3pt]
    Luminosity [$\text{cm}^{-2}~\text{s}^{-1}$]      & $2\times 10^{33}$ & $10^{34}$  & $10^{36}$ \\[3pt]
     Integrated luminosity [$\text{fb}^{-1}$]      & $60$ & $300$  & $3\times 10^4$ \\[3pt]
\hline\hline
\end{tabular*}
\end{table}

\section{semi-inclusive production mechanism}\label{Sec:general_machanism}

\begin{figure}[tbp]
    \includegraphics[width=0.45\columnwidth]{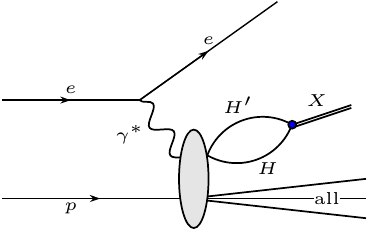}
    \caption{\label{Fig:ep_production}Semi-inclusive electroproduction of the $HH'$ hadronic molecule $X$ in $ep$ collisions. The
    other particles produced in this process are generally denoted by ``$\text{all}.$''} 
\end{figure}

As shown in Fig.~\ref{Fig:ep_production}, the mechanism for the semi-inclusive electroproduction of a hadronic molecule involves a virtual photon, radiated by the electron, interacting with the proton to produce the hadron pair $HH'$ and other particles denoted by ``$\text{all}$''~\cite{Yang:2021jof}; then, the $HH'$ pair is bound together to form a hadronic molecule through the FSI. Therefore, the amplitude of the semi-inclusive production of the hadronic molecule $X$, can be factorized into a short-distance part and a long-distance part~\cite{Braaten:2005jj,Artoisenet:2009wk,Guo:2014ppa,Guo:2014sca},\footnote{In the framework used here, the long-distance part refers to the formation of hadronic molecules from the hadron-pair interactions, and thus the momentum scale is roughly the binding momentum $\kappa=\sqrt{2\mu B}$, where $\mu$ is the reduced mass of the two-body system and $B$ refers to the binding energy.
In this sense, momentum scales that are much larger than $\kappa$  qualify as short-distance scales. The production of charmed hadrons through partonic interactions is thus a short-distance process.} 
\begin{align}
    {\cal M}[X+\all]={\cal M}[HH'+\all]\cdot G \cdot T_X,
    \label{Eq:amp_total}
\end{align}
where $T_X$ denotes the amplitude for the long-distance process of fusing the $HH'$ pair into the hadronic molecule $X$, which can be approximated by the effective coupling constant for $X$ to its constituents in the case of loosely bound hadronic molecules, and the rest is the short-distance part. ${\cal M}[HH'+\all]$ denotes the amplitude for the short-distance process $ep\to HH'+\all$, and $G$ denotes the Green's function of the $HH'$ pair, which is ultraviolet divergent and the divergence can be absorbed by ${\cal M}[HH'+\all]$~\cite{Braaten:2005jj}. 

For the production process $ep\to HH'+\all$, the semi-inclusive production of $HH'$ at short distances may be simulated using MC event generator Pythia~\cite{Sjostrand:2006za}. The most important hard scattering process is $\gamma g\to c\bar{c}$ for the production of charm-hadron pairs considered here. In our explicit
realization, we choose the hard process in Pythia to generate the partonic events and then the $HH'$ events can be obtained after the hadronization. The general differential cross section for the inclusive production of $HH'$ in the MC event generator is
\begin{align}
    d\sigma[HH'(k)]_{\text{MC}}=K_{HH'}\frac{1}{\text{flux}}\sum_{\text{all}}\int d\phi_{HH'+\text{all}}\big|{\cal M}[HH'+\text{all}]\big|^2\frac{d^3k}{(2\pi)^32\mu},
    \label{Eq:cross_MC}
\end{align}
where $k$ and $\mu$ are the three-momentum in the c.m. frame and the reduced mass of the $HH'$ system, respectively. The difference between the MC simulation and the experimental data is amended by an overall factor $K_{HH'}$. To estimate the cross sections for the semi-inclusive electroproduction of hadronic molecules at the order-of-magnitude level, we roughly set $K_{HH'}\simeq 1$. The total cross section for the semi-inclusive production of a hadronic molecule $X$ can be derived from Eqs.~\eqref{Eq:amp_total} and \eqref{Eq:cross_MC}, 
\begin{align}
    \sigma[X+\all]\simeq \frac{{\cal N}}{4m_Hm_{H'}} \left|G T_X\right|^2\left(\frac{d\sigma[HH'+\all]}{dk}\right)_{\text{MC}}\frac{4\pi^2\mu}{k^2},
    \label{Eq:cross_section_ep}
\end{align}
where $m_{H}(m_{H'})$ is the mass of $H(H')$, $\mathcal{N}=2/3$ for the production of the  $\Sigma_c \bar{D}^*$ and $\Xi_c\bar{D}^*$ molecular states with quantum numbers $J^P=3/2^-$ (see Appendix~\ref{Sec:BV_cross_section}), and ${\cal N}=1$ for the other hadronic molecules considered in this work. Here we neglect the scattering between $H(H')$ and other final-state particles. Then, the differential cross section for $HH'$ production in the MC event generator is
\begin{align}
    \left(\frac{d\sigma[HH'+\all]}{dk}\right)_{\text{MC}}\propto k^2.\label{Eq:MC_diff_sec}
\end{align}

The Green's function $G$ in Eq.~\eqref{Eq:cross_section_ep} is ultraviolet divergent and can be regularized by a Gaussian regulator as~\cite{Nieves:2012tt}
\begin{align}
    G(E,\Lambda)=-\frac{\mu}{\pi^2}\left\{\sqrt{2\pi}\frac{\Lambda}{4}+\frac{k\pi}{2}e^{-2k^2/\Lambda^2}\left[i-\text{erfi}\left(\frac{\sqrt{2}k}{\Lambda}\right)\right]\right\},
    \label{Eq:green_function}
\end{align}
where the three-momentum $k$ is related to the energy $E$ of the $HH'$ pair, i.e., $k^2=2\mu(E-m_H-m_{H'})$, and  $\text{erfi}(z)=2/\sqrt{\pi}\int_0^z e^{t^2}dt$ is the imaginary error function. The short-distance amplitude for producing the $HH'$ pair in Eq.~\eqref{Eq:amp_total} should scale as $\Lambda^{-1}$ so as to absorb the leading $\Lambda$ dependence in Eq.~\eqref{Eq:green_function}~\cite{Braaten:2005jj}. Here, since we have taken the short-distance amplitude to be fixed from the MC event generator, which does not have a $\Lambda$ dependence, we choose the cutoff $\Lambda$ to be in the reasonable range $0.5-1.0$ GeV as an estimate~\cite{Guo:2013sya,Liu:2019tjn,Yang:2020nrt,Dong:2021juy}. 

For all of the considered hadronic molecules, the mass is close to the threshold of  $HH'$ which couples to the hadronic molecule in an $S$ wave. Thus, the amplitude $T_X$ in Eq.~\eqref{Eq:cross_section_ep} can be approximated by an effective coupling constant $\geff$.
The coupling constant $\geff$ can be extracted from the $T$ matrix for the low-energy scattering process $HH' \to HH'$, 
\begin{align}
    \geff^2 
    =\lim_{E\to E_{0}} (E^2-E^2_{0})T(E),
    \label{Eq:coupling}
\end{align}
where $E_0$ is the pole position in the complex $E$ plane of the $HH'$ scattering $T$ matrix. We have $E_{0}=M_X$ for a bound or virtual state (with the pole on the first or second Riemann sheet) and $E_0=M_X-i\Gamma/2$ for a resonance (with the pole on the second Riemann sheet). 
Here $M_X$ and $\Gamma$ are the mass and decay width of the hadronic molecule $X$, respectively.
Near the threshold, we only consider a constant contact term as the interaction kernel $V$ for the scattering process $HH'\to HH'$. Then, the $T$ matrix for the scattering of $HH' \to HH'$ can be calculated by solving the Lippmann-Schwinger equation,
\begin{align}
    T(E)=\frac{V}{1-V G(E,\Lambda)}.
    \label{Eq:T_matrix}
\end{align}
Poles of $T$ matrix on the first or second Riemann sheet of the complex $E$ plane satisfy the equation $\text{det}[1-VG(E_0,\Lambda)]=0$. Note that in the evaluation of the effective coupling, Eq.~\eqref{Eq:coupling}, $T(E)$ should take its value according to the specific Riemann sheet where the pole is located.

We consider isospin-breaking effects for the $X(3872)$ by considering both the $D^0\bar{D}^{0*}+\text{c.c.}$ and $D^+D^{*-}+\text{c.c.}$ channels, and the $T$ matrix is
\begin{align}
    V=\frac{1}{2}
    \begin{pmatrix}
      C_0+C_1 & C_0-C_1\\
      C_0-C_1 & C_0+C_1
    \end{pmatrix},
\end{align}
where $C_{0}$ and $C_{1}$ are the low-energy constants for the isoscalar and isovector $D\bar D^*+\text{c.c.}$ channels, respectively. The two low-energy constants can be solved using the $X(3872)$ mass~\cite{LHCb:2020xds} and the isospin-violation ratio~\cite{Hanhart:2011tn} for the $X(3872)$ decays to $J/\psi \pi^+\pi^-$ and $J/\psi \pi^+\pi^0\pi^-$ as inputs~\cite{Hidalgo-Duque:2012rqv}. 
We also take the contact term as the potential for $Z_c(3900)^0$ and $Z_c(3900)^+$, where the values of the low-energy constants are fixed in Ref.~\cite{Yang:2020nrt} from fitting to the BESIII data on $Z_{cs}(3985)$~\cite{BESIII:2020qkh}.
For the other exotic states discussed in this work, we neglect the isospin-breaking effects and extract the coupling constants from the single channel $T$ matrix. 

\begin{figure}[tbp]
    \includegraphics[width=0.45\columnwidth]{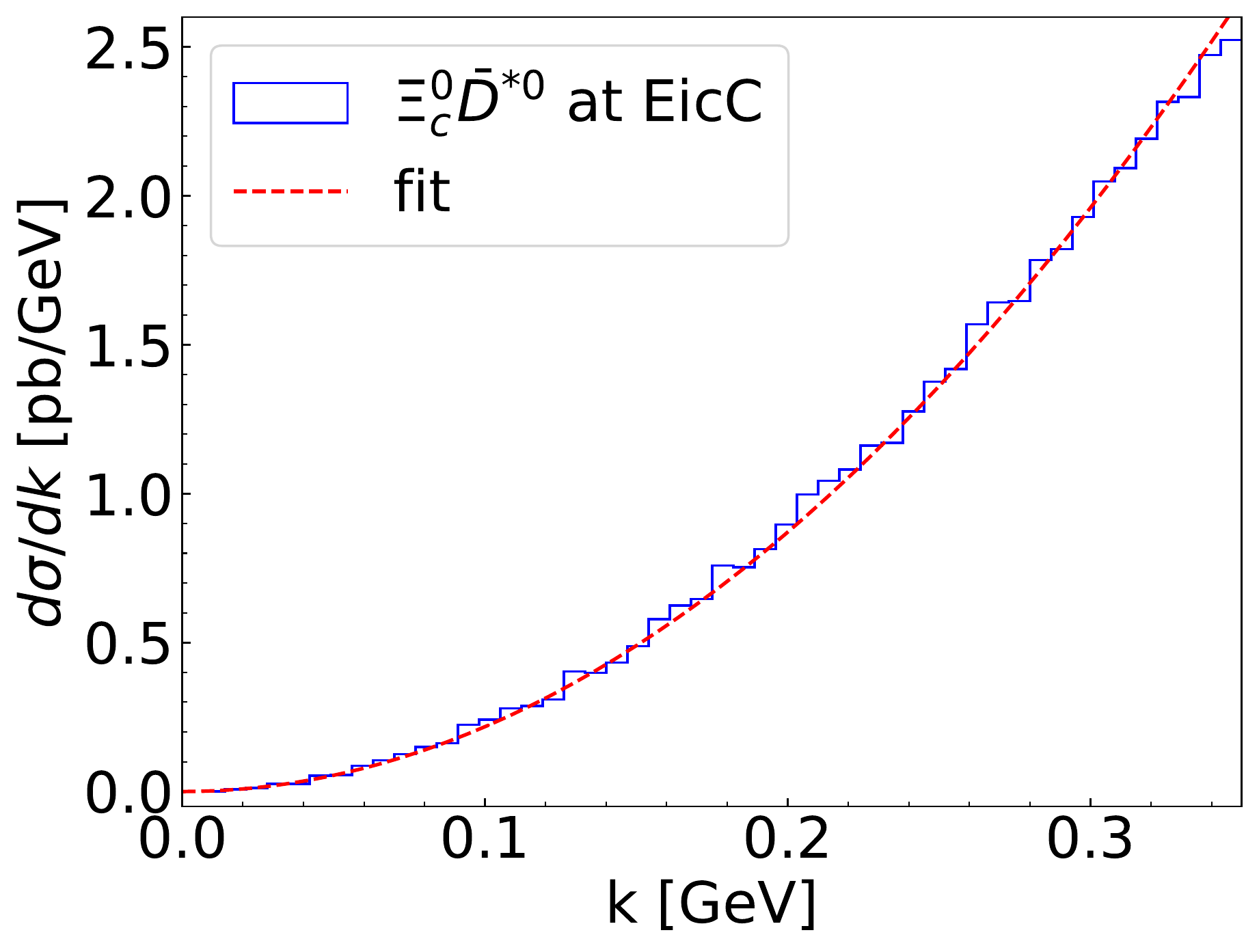} 
    \includegraphics[width=0.45\columnwidth]{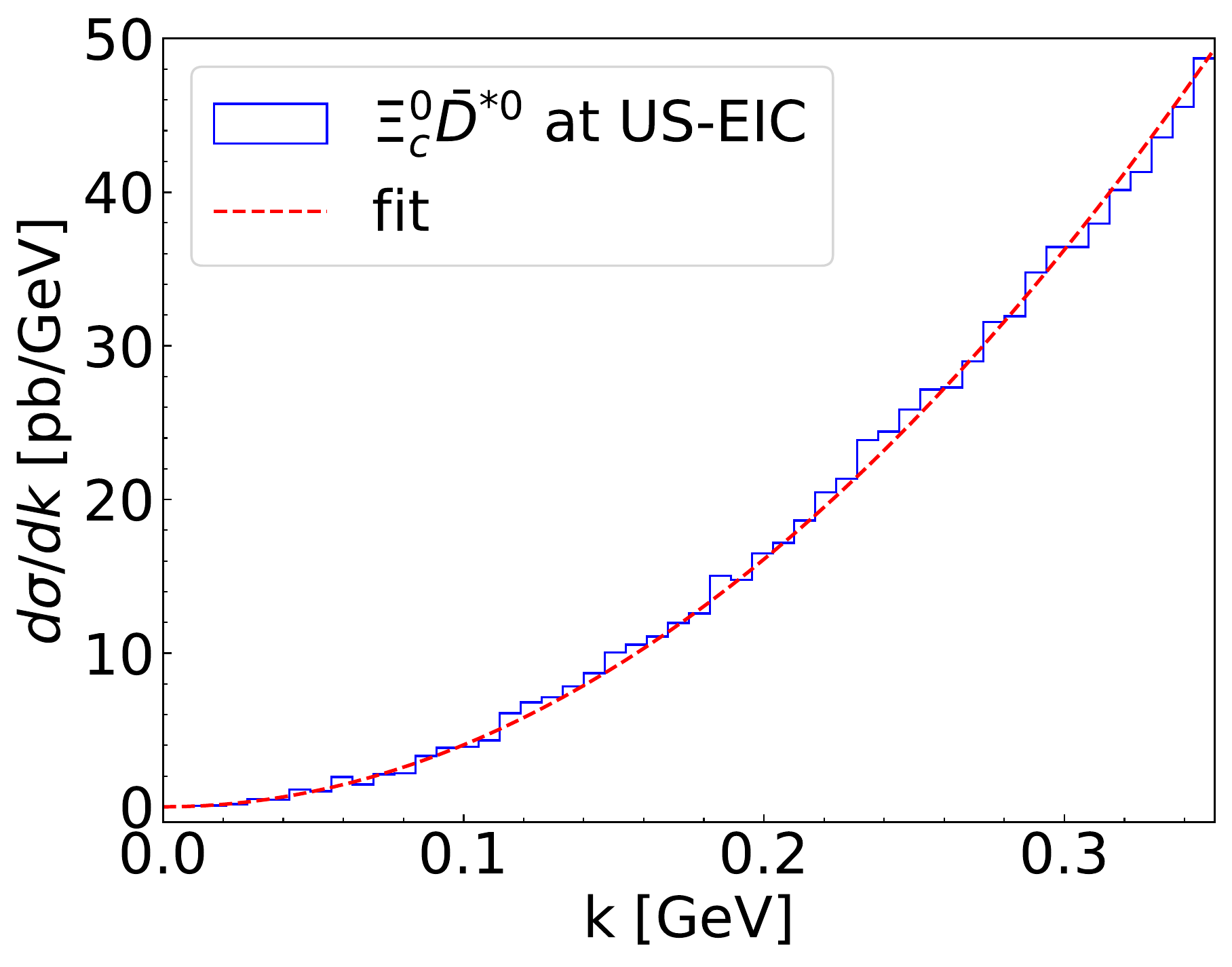}
    \includegraphics[width=0.45\columnwidth]{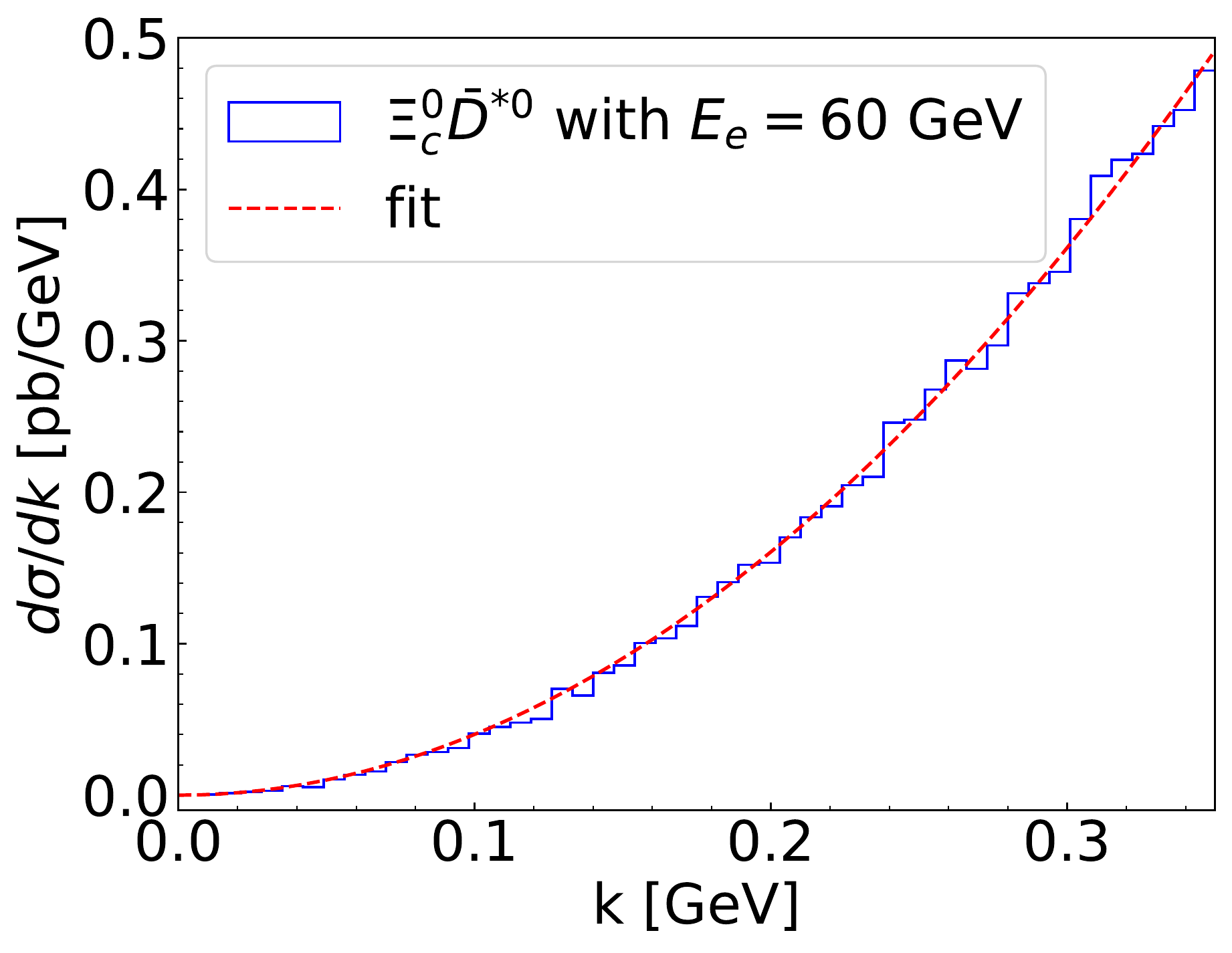} 
    \includegraphics[width=0.45\columnwidth]{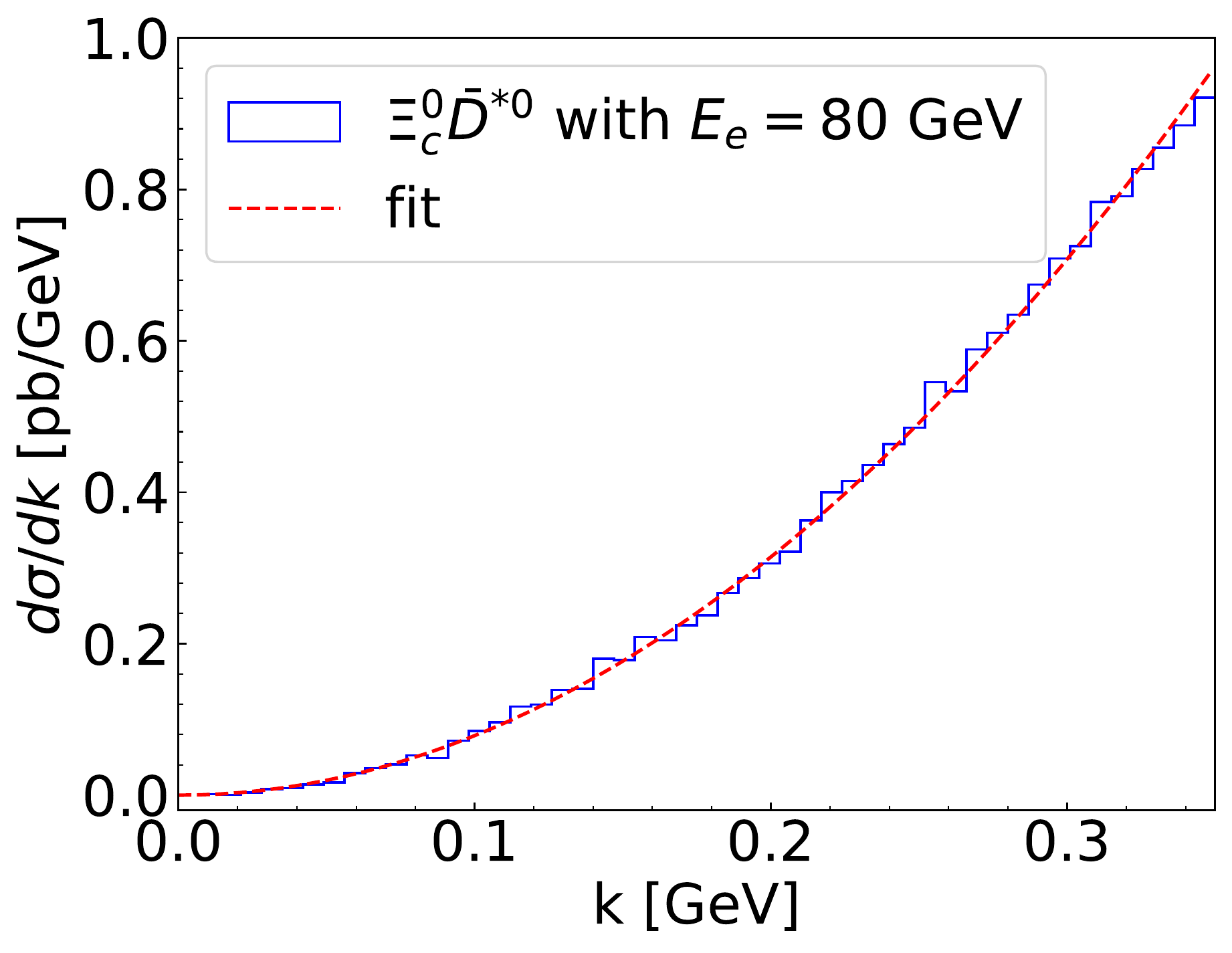}
    \includegraphics[width=0.45\columnwidth]{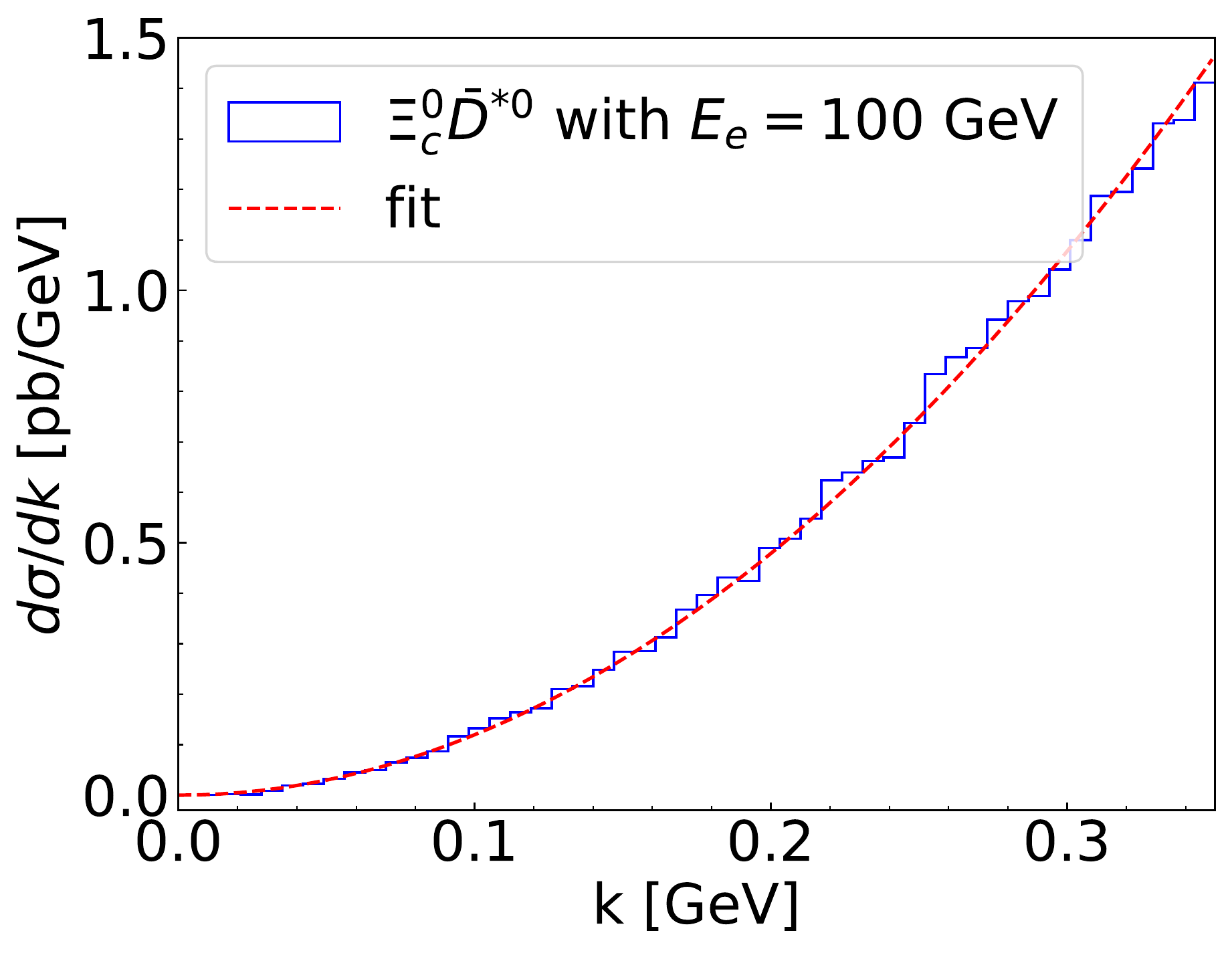} 
    \includegraphics[width=0.45\columnwidth]{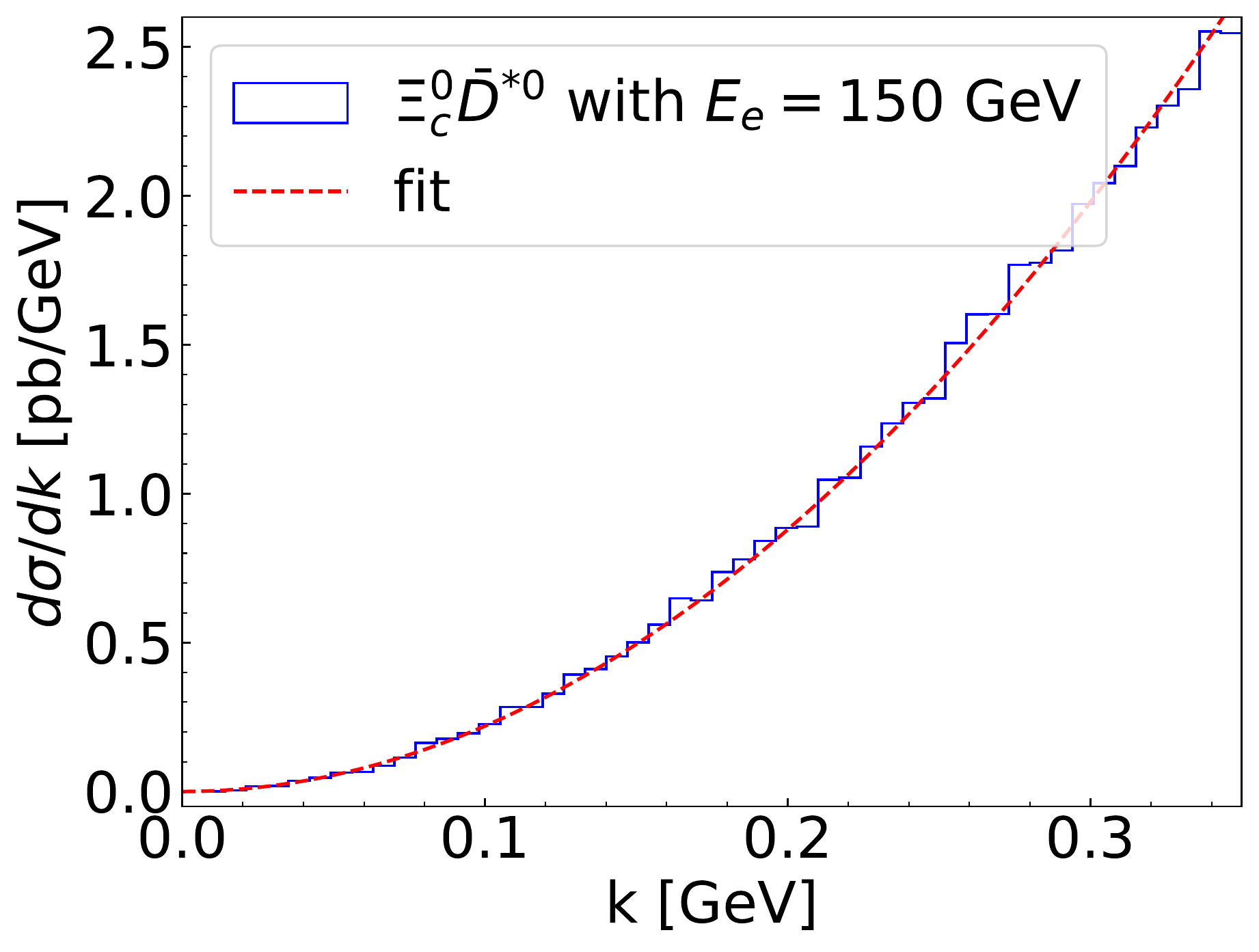}
    \caption{Differential cross sections d$\sigma/$d$k$ for the production of $\Xi_c^0\bar{D}^{*0}$ in $ep$ collisions. The histograms are generated using Pythia and the dashed curves are from fits using d$\sigma/$d$k\propto k^2$. In the lower four plots, the proton is at rest and $E_e$ denotes the energy of the electron beam. }
    \label{Fig:diff_sec_Xi_D1}
\end{figure}

\section{Numerical results}\label{Sec:result}

\begin{table}[tb]
\caption{\label{Tab:Sec_EicC_EIC} Order-of-magnitude estimates of the cross sections (in units of pb) for $ep\to X+\text{all}$ at EicC and EIC, where $X$ denotes one of the two $T_{cc}$ states, three $P_{cs}$ states, and some hidden-charm baryon-antibaryon molecules. The quantum numbers for those states are listed in the third column. The binding energies, defined as $m_1+m_2-M$ where $m_{1,2}$ are the masses of the constituents and $M$ is the mass of the hadronic molecule, are listed in the fourth column. The results outside (inside) parentheses denote the cross sections with $\Lambda=0.5$ GeV ($1.0$ GeV). }
\renewcommand{\arraystretch}{1.2}
\begin{tabular*}{\columnwidth}{@{\extracolsep\fill}lccccc}
\hline\hline                                           
           & Constituents  & $IJ^{P(C)}$   & Binding energy [MeV]  & EicC [pb] & EIC [pb]\\[3pt]
\hline
     $T_{cc}^+$      & $DD^*$ & $01^+$ & 0.273~\cite{LHCb:2021auc,LHCb:2021vvq}    & $0.3\times 10^{-3}~(1.2\times 10^{-3}) $     & 0.1 (0.5)  \\[3pt]
     $T_{cc}^*$  & $D^*D^*$ & $01^+$ & $0.503$~\cite{Du:2021zzh}    & $0.2\times 10^{-3}~(1.0\times 10^{-3}) $    & $0.1~ (0.4)$ \\[3pt]
    $P_{cs}$  & $\Xi_c{\bar D}$ & $0\frac{1}{2}^-$  & $0.3 ~(3.53)$~\cite{Dong:2021juy}     & $0.1 ~(1.6)$   & $1.8 ~(30)$  \\[3pt]
    $P_{cs}$  & $\Xi_c{\bar D}^*$ & $0\frac{1}{2}^-$  & $18.83$~\cite{Du:2021bgb}    & $0.1 ~(0.5)$  & $1.3~ (8.8)$  \\[3pt]
    $P_{cs}$  & $\Xi_c{\bar D}^*$ & $0\frac{3}{2}^-$  & $18.83$~\cite{Du:2021bgb}    & $0.1~ (0.9)$   & $2.6~ (18)$  \\[3pt]
    & $\Lambda_c {\bar \Lambda}_c$ & $00^{-+}$  & $1.98 ~(33.8)$~\cite{Dong:2021juy}     & $0.3~ (3.0)$     & $9.6~ (110)$  \\[3pt]
    & $\Sigma_c {\bar \Sigma}_c$ & $00^{-+}$  & $11.1 ~(60.8)$~\cite{Dong:2021juy}      & $0.7\times 10^{-3}~(5.2\times 10^{-3}) $    & $0.04~(0.29) $ \\[3pt]
    & $\Sigma_c {\bar \Sigma}_c$ & $10^{-+}$  & 8.$28 ~(53)$~\cite{Dong:2021juy}      &  $0.7\times 10^{-3}~(5.3\times 10^{-3}) $    & $0.04~(0.29) $ \\[3pt]
   & $\Xi_c {\bar \Xi}_c$ & $00^{-+}$  & $4.72~ (42.2)$~\cite{Dong:2021juy}     & $1.4\times 10^{-3}~(1.1\times 10^{-2}) $    & 0.1 (0.5) \\[3pt]
   & $\Xi_c {\bar \Xi}_c$ & $10^{-+}$  & $18.2~ (0.39)$~\cite{Dong:2021juy}     & $0.1\times 10^{-3}~(1.7\times 10^{-3}) $    & $3.9\times 10^{-3}~(7.1\times 10^{-2})$ \\[3pt]
   & $\Lambda_c {\bar \Sigma}_c$ & $10^{-}$  & $2.19~ (33.9)$~\cite{Dong:2021juy}     & $0.01~(0.12) $    & $0.5~ (5.5)$ \\[3pt]
   & $\Lambda_c {\bar \Xi}_c$ & $\frac{1}{2}0^{-}$  & $1.29~ (8.42)$~\cite{Dong:2021juy}    & $0.01~(0.14) $    & $0.2 ~(5.3)$ \\[3pt]
   & $\Xi_c {\bar \Sigma}_c$ & $\frac{1}{2}0^{-}$  & $5.98~ (46.4)$~\cite{Dong:2021juy}    & $0.8\times 10^{-3}~(7.3\times 10^{-3}) $    & $0.04~(0.36) $  \\[3pt]
\hline\hline
\end{tabular*}
\end{table}

The production mechanism discussed in Sec.~\ref{Sec:general_machanism} is applied to estimate the cross sections for the semi-inclusive electroproduction of hadronic molecules. We simulate the $ep$ collisions at EicC, EIC, and CEBAF (24~GeV) with the MC event generator Pythia~\cite{Sjostrand:2006za} and get the differential cross sections for the semi-inclusive production of the $HH'$ pair, $\left(d\sigma[HH'+\all]/dk\right)_{\text{MC}}$. As discussed in Eq.~\eqref{Eq:MC_diff_sec}, the differential cross section for producing the $HH'$ pair is proportional to $k^2$ in the small-momentum region (in this work, we choose $|k|< 350$~MeV). For instance, the differential MC cross sections for $\Xi_c^0\bar{D}^{*0}$ pair production at EicC, EIC and in fixed-target $ep$ collisions,\footnote{The c.m. collision energy for the 24~GeV upgrade of CEBAF~\cite{Arrington:2021alx} is lower than 10~GeV, the lower limit for the use of Pythia~\cite{Sjostrand:2006za}. To estimate the cross sections, we calculate the cross sections at the electron energies $E_e= 60, 80, 100,$ and 150~GeV, and then extrapolate the results to $E_e=24$~GeV.} as shown in Fig.~\ref{Fig:diff_sec_Xi_D1}, are exactly proportional to $k^2$. 

For the long-distance part, as discussed above, we use the cutoff-dependence formula in Eq. \eqref{Eq:cross_section_ep} with the cutoff $\Lambda \in [0.5,1.0]$~GeV for an order-of-magnitude estimate of the cross section.
To extract the effective coupling $\geff$, the masses and quantum numbers of the hadronic molecules in Tables~\ref{Tab:Sec_EicC_EIC} and~\ref{Tab:Sec_JLab} are fixed to experimental observations, whenever possible, and theoretical predictions.
For the $T_{cc}$ states, the mass of $T_{cc}^+$ is set by the central value  reported by the LHCb Collaboration~\cite{LHCb:2021auc,LHCb:2021vvq}, while the $T_{cc}^*$ mass is set to the theoretical prediction in Ref.~\cite{Du:2021zzh}. $P_{cs}(4459)$, which was observed by the LHCb Collaboration~\cite{LHCb:2020jpq}, might be a $\Xi_c\bar{D}^*$ bound state with $J^P=1/2^-$ or $3/2^-$ or the signal of both states~\cite{Liu:2020hcv,Chen:2020kco,Peng:2020hql,Chen:2020uif,Dong:2021juy}, and thus we estimate its production rates with both $J^P=1/2^-$ and $3/2^-$. In Ref.~\cite{Dong:2021juy}, more than 200 hidden-charm hadronic molecules were predicted. In addition to those whose production rates have been estimated in Ref.~\cite{Yang:2021jof}, we consider here the $\Xi_c \bar{D}$ molecule and a set of hidden-charm baryon-antibaryon molecules, composed of $\Lambda_c\bar{\Lambda}_c$, $\Sigma_c\bar{\Sigma}_c$, $\Xi_c\bar{\Xi}_c$, $\Lambda_c \bar{\Sigma}_c$, $\Lambda_c\bar{\Xi}_c$ and $\Xi_c\bar{\Sigma}_c$. Among these states predicted in Ref.~\cite{Dong:2021juy}, the $\Xi_c\bar{\Xi}_c$ molecule with $IJ^{PC}=10^{-+}$ is a virtual state; the $\Xi_c \bar{D}$ and $\Lambda_c \bar{\Xi}_c$ molecules are virtual states when $\Lambda=0.5$~GeV, while they are bound states when $\Lambda=1$~GeV; the other states are bound states.
Following Refs.~\cite{Sakai:2019qph,Du:2019pij,Du:2021fmf}, we assign $P_c(4312)$ and the narrow $P_c(4380)$~\cite{Du:2019pij} as the $\Sigma_c\bar{D}$ molecule with $J^P=1/2^-$ and the $\Sigma_c^* \bar{D}$ molecule with $J^P=1/2^-$, respectively, and $P_c(4440)$ and $P_c(4457)$ as the $\Sigma_c \bar{D}^*$ molecules with $J^P=3/2^-$ and $1/2^-$. Following Ref.~\cite{Yang:2021jof}, we use the $P_c$ masses from scheme II in Ref.~\cite{Du:2019pij} as inputs to extract the effective coupling constants $\geff$.

We estimate the cross sections for the production of the molecules, as shown in Table~\ref{Tab:Sec_EicC_EIC}, at EicC and EIC with the energy configurations in Table~\ref{Tab:Lum_EicC_EIC}. Our results show that the cross sections at EIC are roughly 1 order of magnitude larger than that at EicC for producing hidden-charm hadronic molecules, which are in line with the estimates in Ref.~\cite{Yang:2021jof}. However, as shown in Table~\ref{Tab:Sec_EicC_EIC}, this difference increases to 2 orders of magnitude for the $T_{cc}$ states. Such a difference is caused by the different charm numbers between the $T_{cc}$ states and hidden-charm states. 
In the processes for the production of the $T_{cc}$ states, at least two charm quarks and two anticharm quarks should be generated in $ep$ collisions. Comparing with the production of the hidden-charm exotic states, higher c.m. collision energies are required to efficiently produce the $T_{cc}$ states. With the energy configurations listed in Table~\ref{Tab:Lum_EicC_EIC}, the c.m. collision energy for EicC is about 17~GeV, which is smaller than that of EIC (about 141 GeV), so the cross sections for producing the $T_{cc}$ states are reasonably small.

As listed in Table~\ref{Tab:Sec_EicC_EIC}, the cross sections for the double-charm states, $T_{cc}^+$ and $T_{cc}^*$, are at the level of 1~fb and 0.1~pb for EicC and EIC, respectively. Thus, the estimated numbers of $T_{cc}^+$ events are $18- 72$ at EicC and $3\times 10^4-1.5\times 10^5$ at EIC with the integral luminosities listed in Table~\ref{Tab:Lum_EicC_EIC}. In particular, our estimate for the $T_{cc}^+$ event number is significantly smaller than the number of double-charm baryon $\Xi_{cc}$ events (about $4.0\times 10^5$ in a year) estimated in Ref.~\cite{Anderle:2021wcy}. Given that the event numbers of $T_{cc}^+$~\cite{LHCb:2021vvq} and $\Xi_{cc}^+$~\cite{LHCb:2017iph} observed by the LHCb Collaboration are of the same order of magnitude,\footnote{The event number for $\Xi_{c c}^{++} \rightarrow \Lambda_{c}^{+}\left(\rightarrow p K^{-} \pi^{+}\right) K^{-} \pi^{+} \pi^{+}$ is $313 \pm 33$ with an integrated luminosity of 1.7~fb$^{-1}$~\cite{LHCb:2017iph}; the event number for $\mathrm{T}_{c c}^{+} \rightarrow D^{0}\left(\rightarrow K^{-} \pi^{+}\right) \bar{D}^{0}\left(\rightarrow K^{+} \pi^{-}\right) \pi^{+}$ is $117 \pm 16$ with an integrated luminosity of 9~fb$^{-1}$~\cite{LHCb:2021vvq}.} investigations are needed to understand the large difference.
The cross sections for the $P_{cs}$ states are around 1~pb at EicC and tens of pb at EIC, so that we expect at least $6\times 10^3$ and $3.9\times 10^5$ events for the $P_{cs}$ states produced at EicC and EIC, respectively. The estimated cross sections of the $P_{cs}$ states are of the same order as those of the $P_c$ states estimated in Ref.~\cite{Yang:2021jof}. 

Our results indicate that, at EicC and EIC, the production rates for the $\Lambda_c\bar{\Lambda}_c$ molecule with quantum numbers $IJ^{PC}=00^{-+}$ are significantly larger than those for the other baryon-antibaryon molecules listed in Table~\ref{Tab:Sec_EicC_EIC}.
The events for the $\Lambda_c\bar{\Lambda}_c$ molecule are about $2\times 10^4 - 2\times 10^5$ and $3 \times 10^6 - 3\times 10^7$ at EicC and EIC, respectively. Therefore, it is promising to find the $\Lambda_c\bar{\Lambda}_c$ molecule and study its properties in detail at EicC and EIC. Besides, for the $\Xi_c\bar{\Xi}_c$ molecule with $IJ^{PC}=10^{-+}$, we expect that about $10 - 100$ and $1\times 10^3 - 2\times 10^4$ events can be produced at EicC and EIC, respectively, and thus it can be searched for at EIC.

\begin{figure}[tbh]
    \includegraphics[height=0.32\columnwidth]{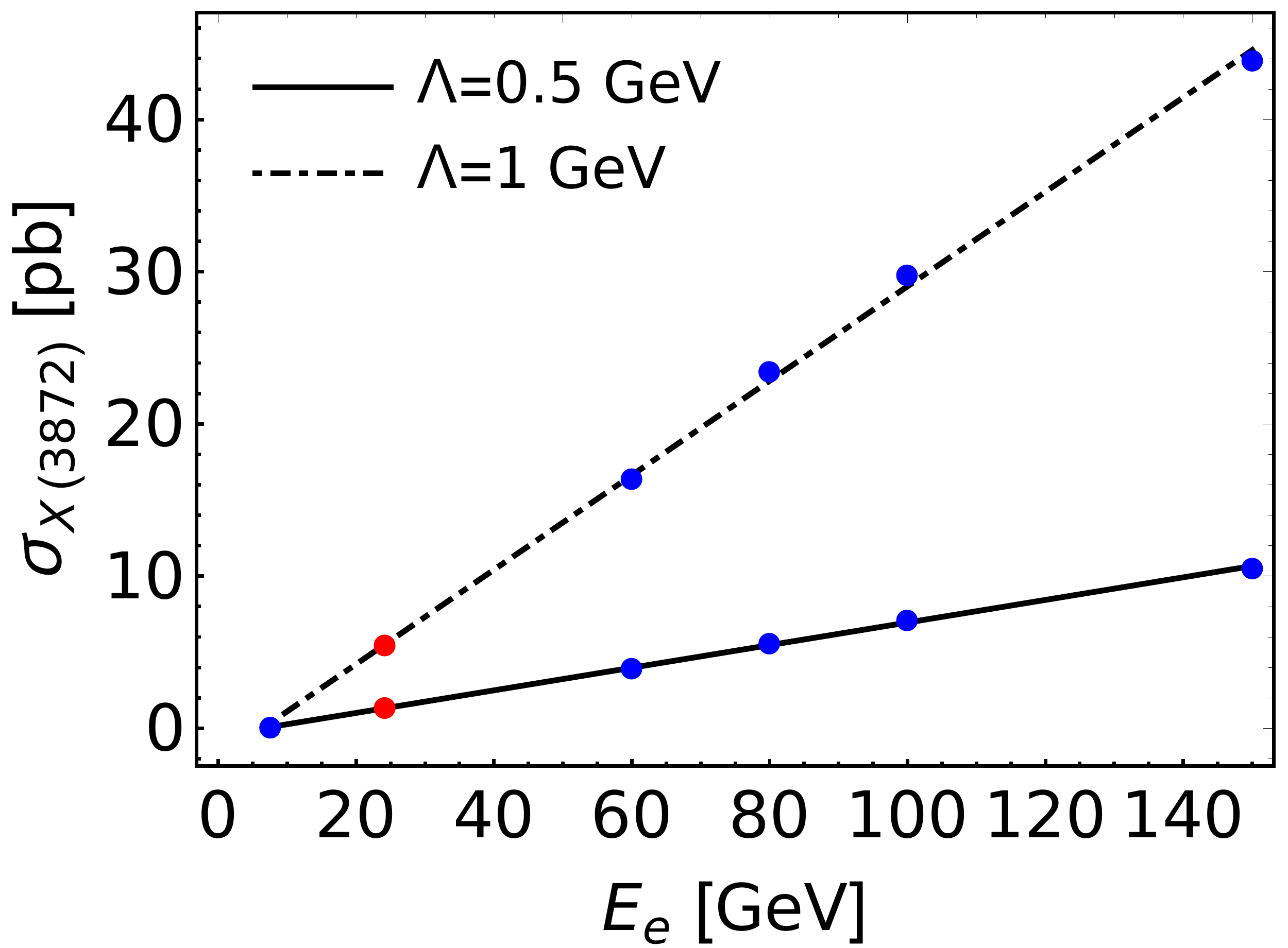} 
    \includegraphics[height=0.32\columnwidth]{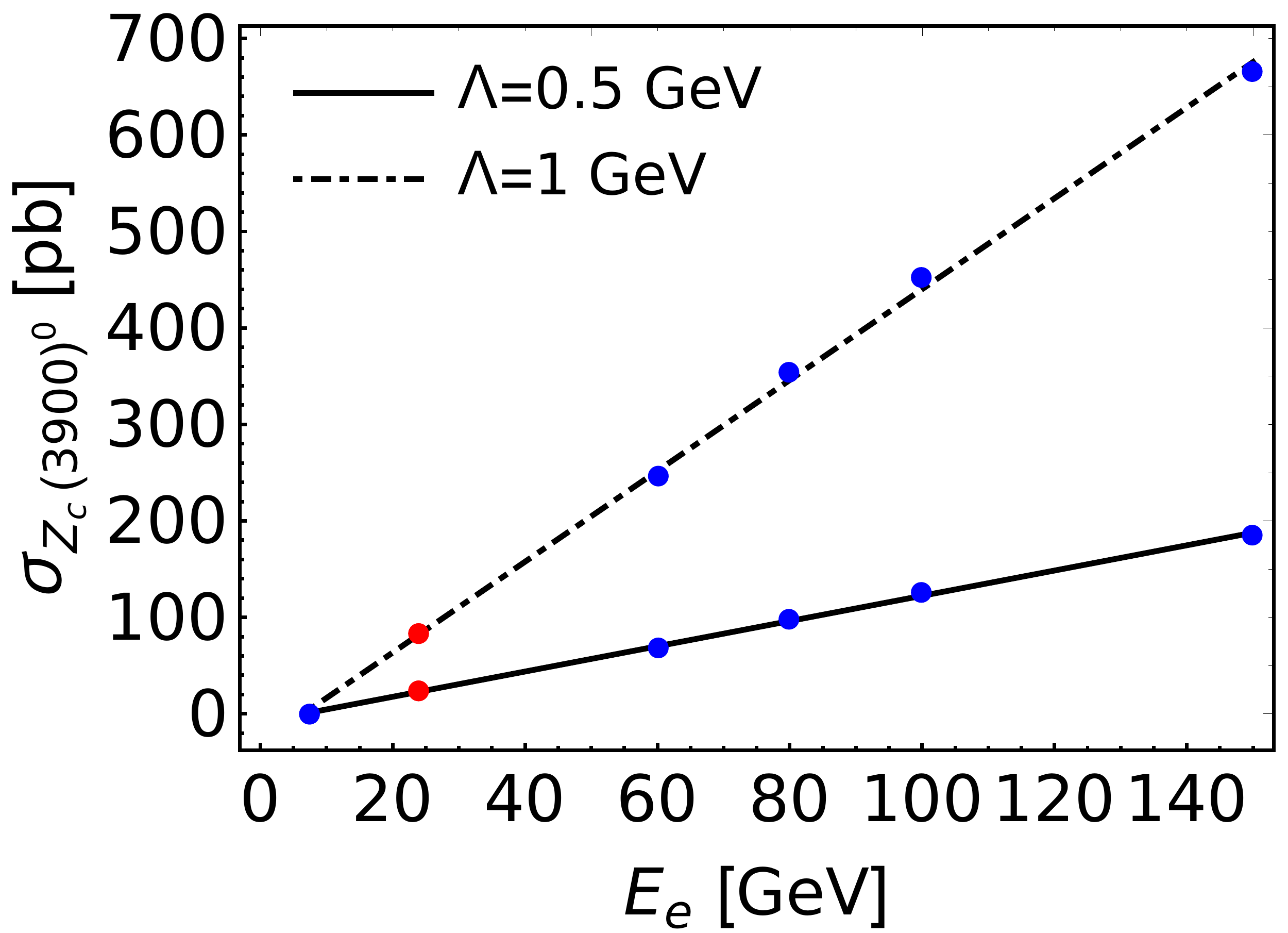}
    \includegraphics[height=0.32\columnwidth]{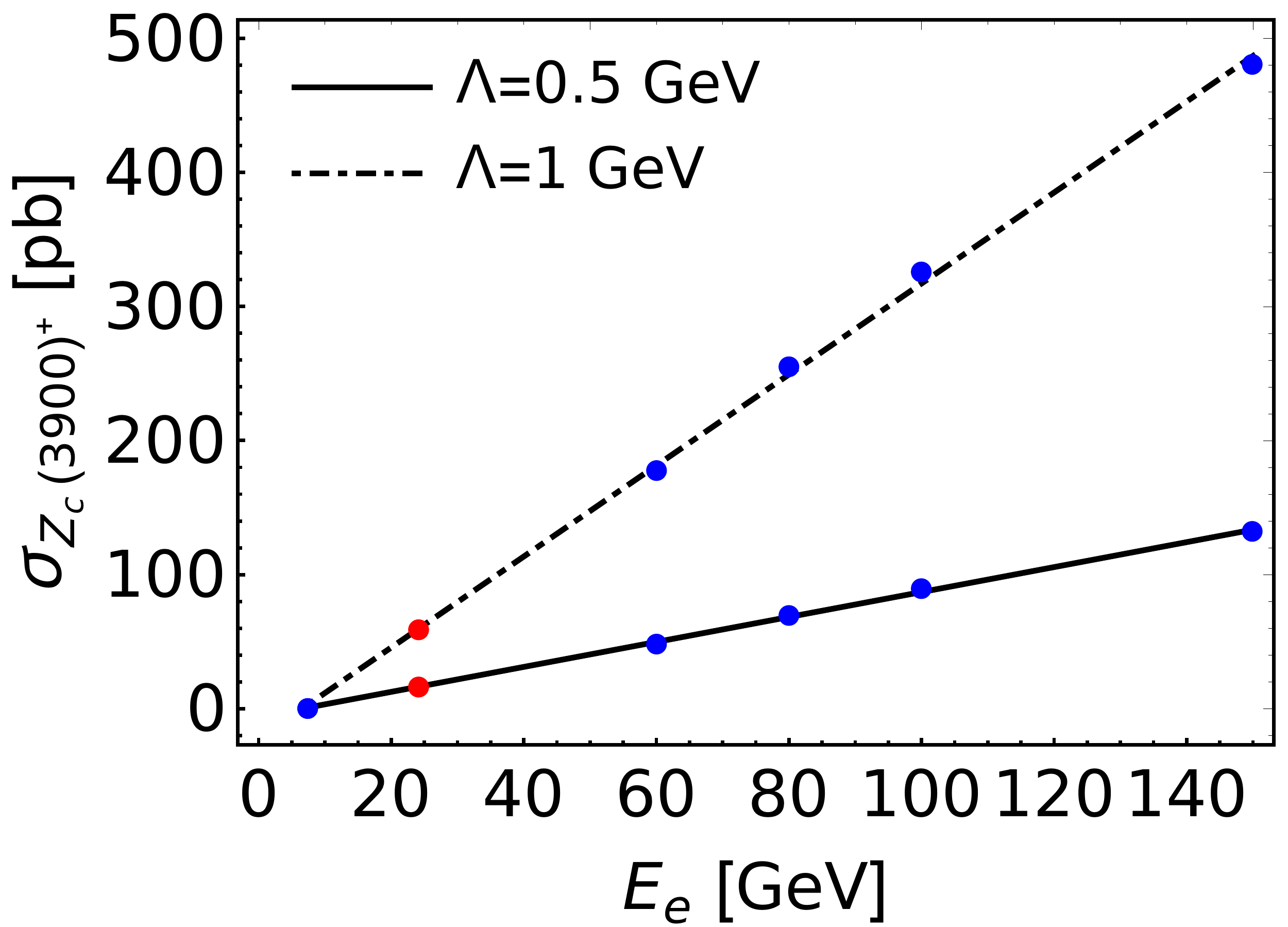} 
    \caption{Semi-inclusive cross sections for producing $X(3872)$ and $Z_c(3900)^{0(+)}$ through the $ep\to X+\text{all}$ reactions with the proton at rest. $E_e$ denotes the electron energy in the process, and $X$ denotes $X(3872)$ or $Z_c$. The blue points represent the cross sections 
    estimated at different electron energies and the red points are the cross sections extrapolated to $E_e=24$~GeV. }
    \label{Fig:JLab_meson_meson}
\end{figure}

\begin{figure}[tbh]
    \includegraphics[height=0.32\columnwidth]{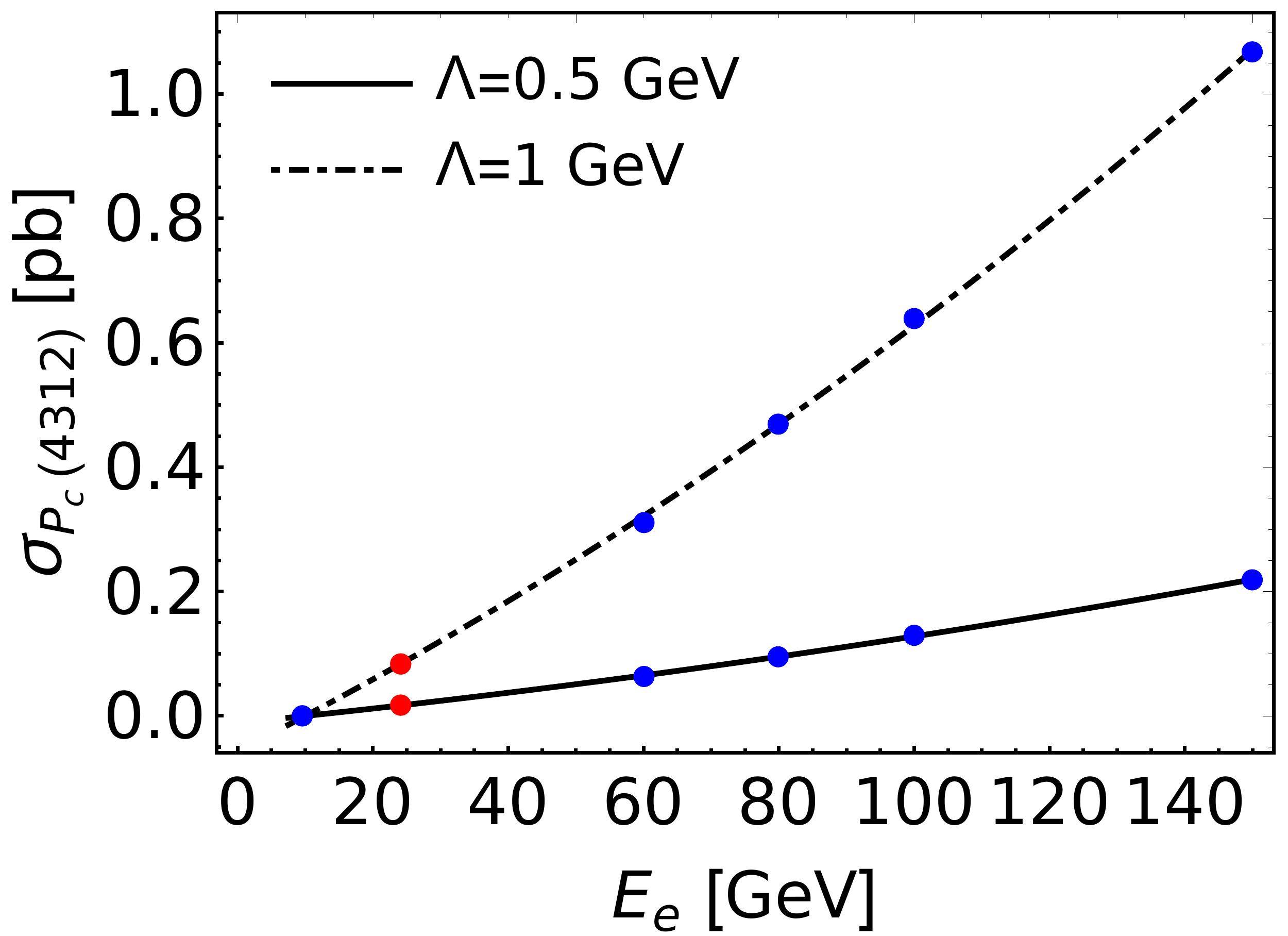} 
    \includegraphics[height=0.32\columnwidth]{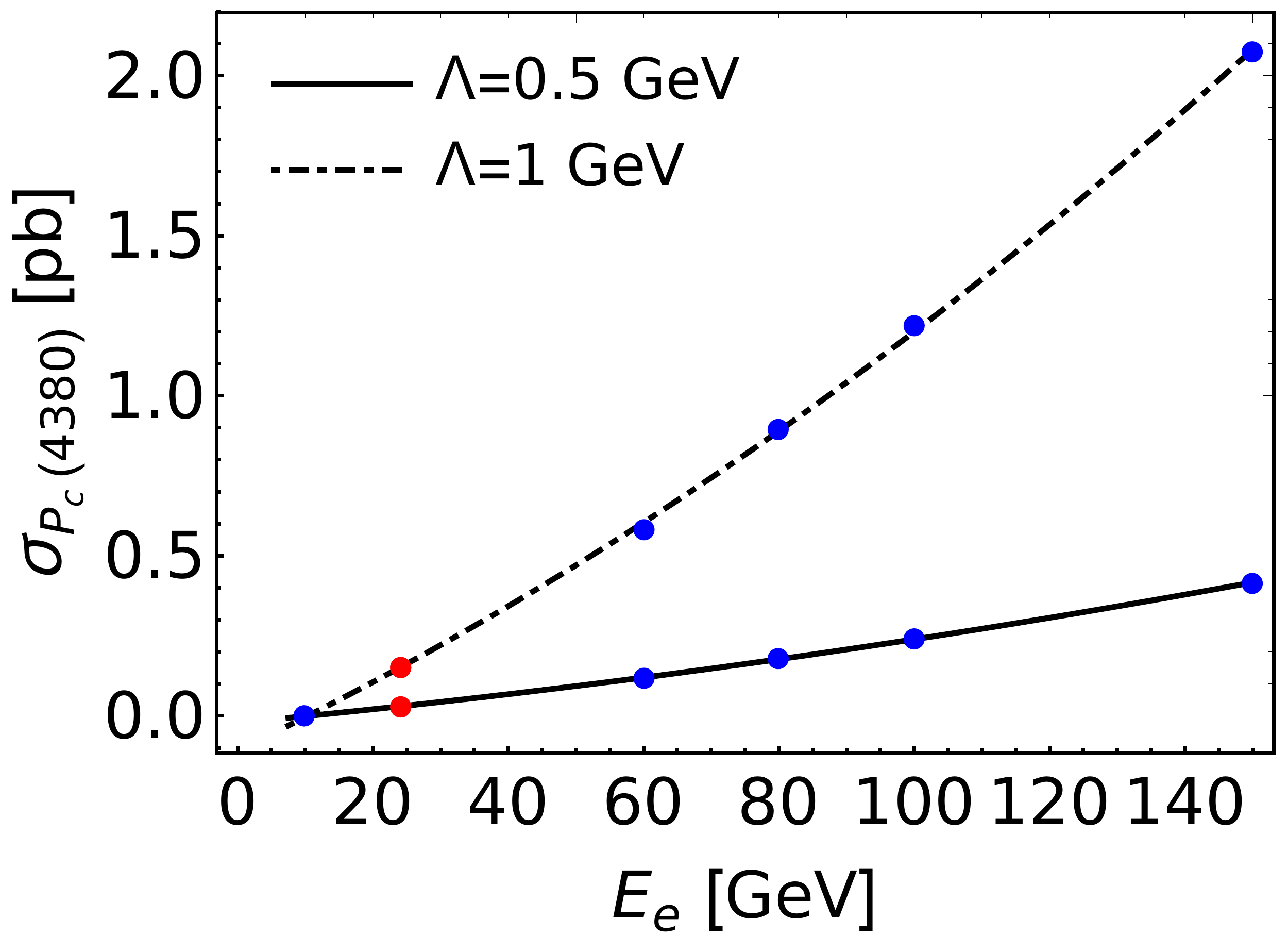}
    \includegraphics[height=0.32\columnwidth]{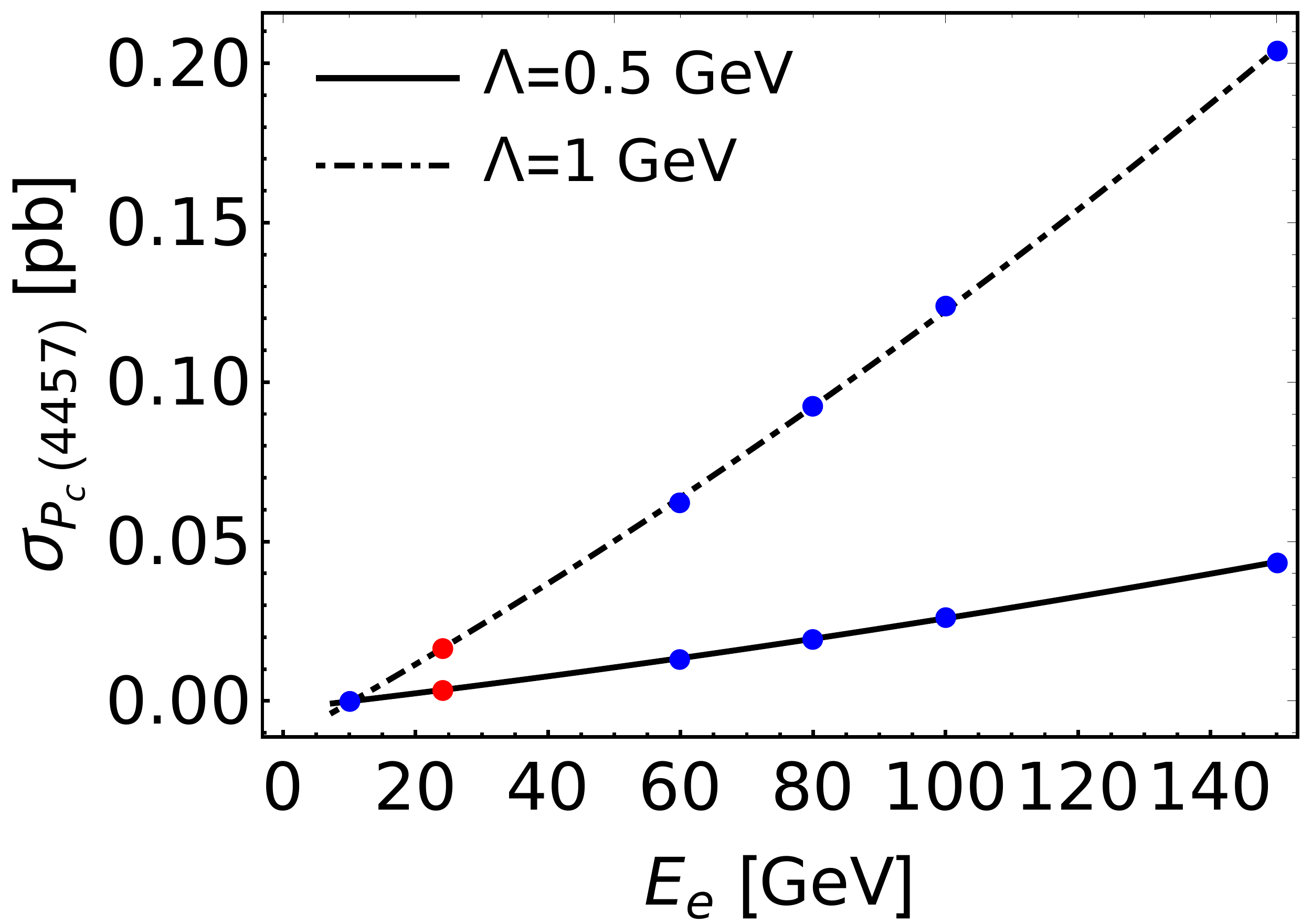} 
    \includegraphics[height=0.32\columnwidth]{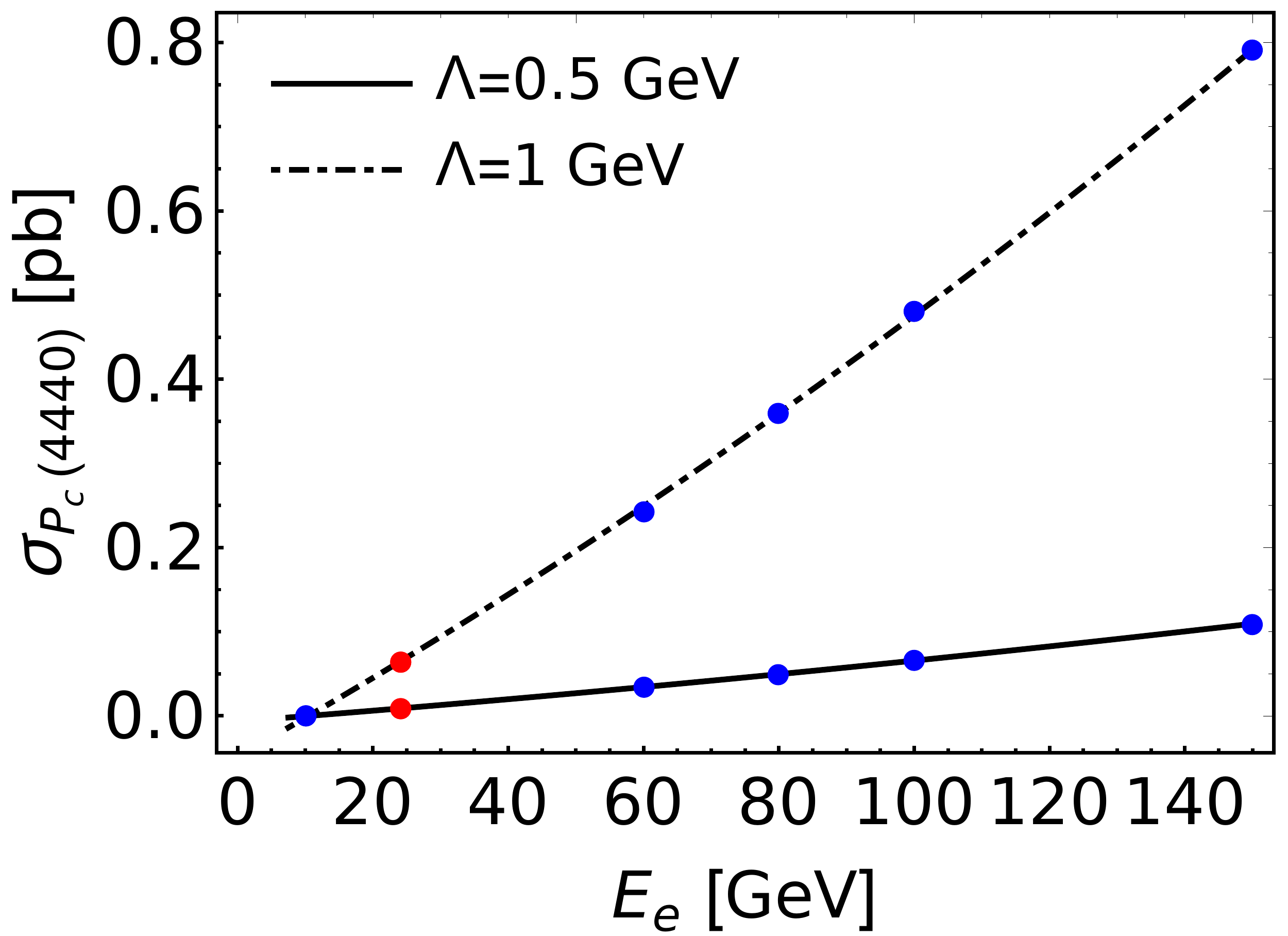}
    \includegraphics[height=0.32\columnwidth]{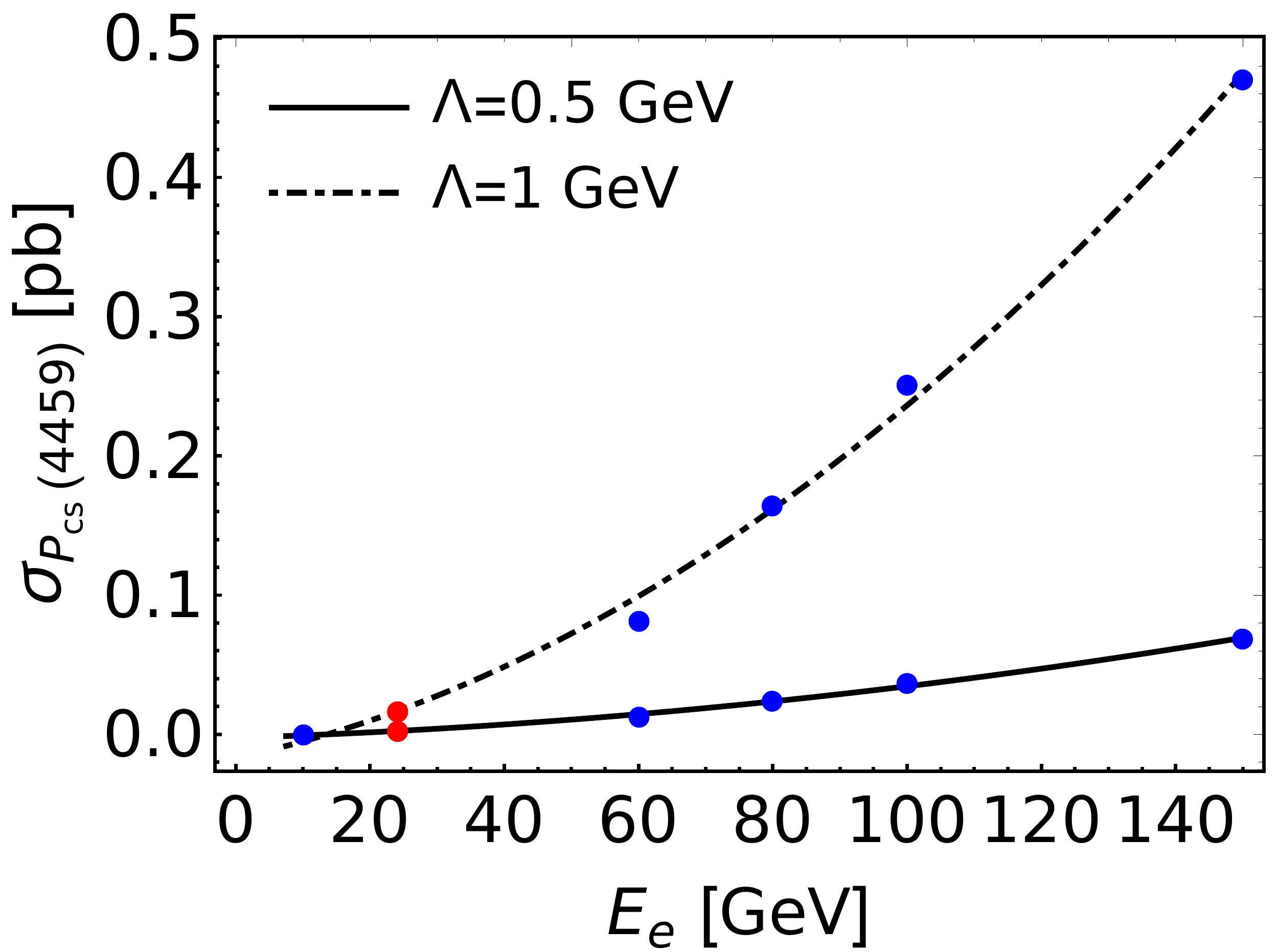} 
    \includegraphics[height=0.32\columnwidth]{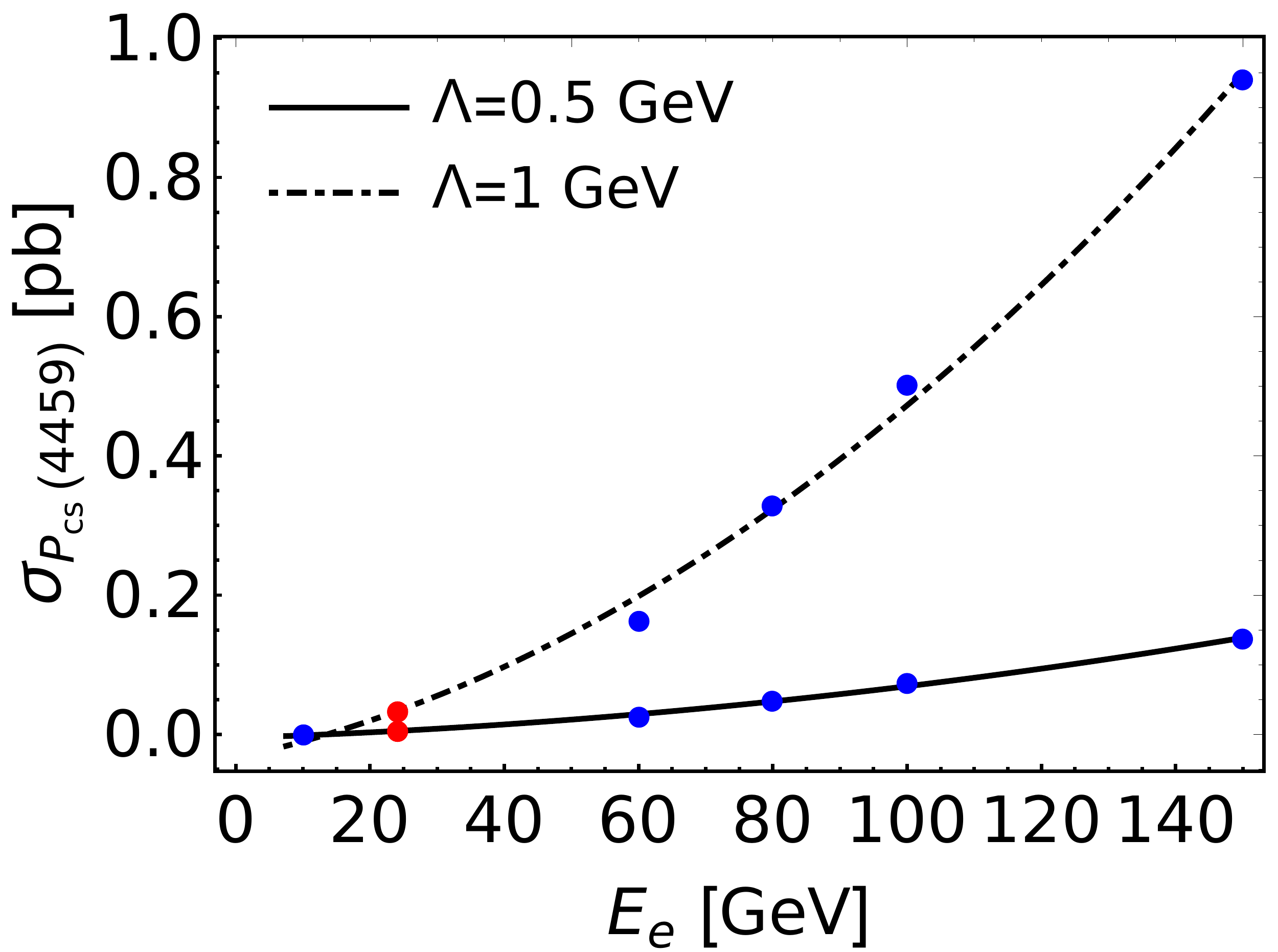}
    \caption{Semi-inclusive cross sections for producing hidden-charm pentaquarks through the $ep\to X+\text{all}$ reactions with the proton at rest. $E_e$ denotes the electron energy in the process, and $X$ denotes $P_c$ or $P_{cs}$. The blue points represent the cross sections
    estimated at different electron energies and the red points are the cross sections extrapolated to $E_e=24$~GeV. In the last row, the left and right plots are the results assuming the spin of the $P_{cs}(4459)$ to be 1/2 and 3/2, respectively.}
    \label{Fig:JLab_meson_baryon}
\end{figure}

As for CEBAF (24~GeV), because the c.m. collision energy is below the applicable energy range of Pythia, we estimate the cross sections for producing the hidden-charm hadronic molecules at higher electron energies (60, 80, 100, and 150~GeV) and then extrapolate the results to 24~GeV. The extrapolation, as shown in Figs.~\ref{Fig:JLab_meson_meson} and~\ref{Fig:JLab_meson_baryon}, is further constrained by requiring the cross section to vanish at the relevant threshold. 
The so-estimated cross sections for the production of a set of hidden-charm hadronic molecules at the proposed 24 GeV upgrade of CEBAF are listed in Table~\ref{Tab:Sec_JLab}.
The cross sections are around 1~pb for producing $X(3872)$, and 0.01~pb for producing the $P_c$ and $P_{cs}$ states. Although the cross sections at CEBAF (24~GeV) are significantly smaller than those at EicC and EIC, its much higher integrated luminosity listed in Table~\ref{Tab:Lum_EicC_EIC} still permits a large number of events for certain hidden-charm exotic hadrons to be produced. 
For instance, 
$\mathcal{O}(10^7\sim10^8)$ $X(3872)$ can be produced through the semi-inclusive processes.
Considering the branching fractions $\text{Br}(X(3872)\to J/\psi \pi\pi)=(3.8\pm 1.2)\%$ and $\text{Br}(J/\psi \to l^+l^-)=12\%$~\cite{Workman:2022ynf}, the event numbers will be $\mathcal{O}(10^5- 10^6)$ for one year of operation. Besides, the cross sections for the $Z_c$ states are 1 order of magnitude larger than that for $X(3872)$, which are compatible with the estimate at EicC and EIC~\cite{Yang:2021jof}. 

\begin{table}[tb]
\caption{\label{Tab:Sec_JLab}Order-of-magnitude estimates of the cross sections for $ep\to X +\all$ at the proposed 24~GeV upgrade of CEBAF, where $X$ denotes one of the hidden-charm hadronic molecules, $X(3872)$, $Z_c(3900)^{0(+)}$, four $P_{c}$ states and two $P_{cs}$ states.  The results outside (inside) the parentheses are the cross sections estimated with  $\Lambda=0.5$ GeV (1.0 GeV).}
\renewcommand{\arraystretch}{1.2}
\begin{tabular*}{\columnwidth}{@{\extracolsep\fill}lcccc}
\hline\hline                                  %
     & {Constituents}  & {$IJ^{P(C)}$}  & {Binding energy [MeV]}   &  $\sigma_X$ [pb] \\[3pt]
\hline
          $X(3872)$      & $D\bar{D}^*$ & $01^{++}$ & $4.15$    & $1.3~(5.5)$    \\[3pt]
    $Z_c(3900)^0$      & $D\bar{D}^*$ & $11^{+-}$ & $-12.57$  & $22.9~(82.4)$    \\[3pt]
    $Z_c(3900)^+$      & $D^{*+}\bar{D}^0$ & $11^{+}$ & $-13.30$  & $16.2~(59.2)$    \\[3pt]
    $P_c(4312)$      & $\Sigma_c\bar{D}$ & $\frac{1}{2}\frac{1}{2}^{-}$ & $6.68$  & $0.02~(0.08)$        \\[3pt]
    $P_c(4440)^+$      & $\Sigma_c\bar{D}^*$ & $\frac{1}{2}\frac{3}{2}^{-}$ & $21.06$  & $0.01~(0.06)$      \\[3pt]
    $P_c(4457)^+$      & $\Sigma_c\bar{D}^*$ & $\frac{1}{2}\frac{1}{2}^{-}$ & $3.06$  & $3.4\times 10^{-3}~(16.4\times 10^{-3})$      \\[3pt]
    $P_c(4380)^+$      & $\Sigma_c^*\bar{D}$ & $\frac{1}{2}\frac{3}{2}^{-}$ & $7.18$  & $0.03~(0.15)$     \\[3pt]
    $P_{cs}(4459)$      & $\Xi_c\bar{D}^*$ & $0\frac{3}{2}^{-}$ & $18.83$ & $4.9\times 10^{-3}~(33.2\times 10^{-3})$     \\[3pt]
    $P_{cs}(4459)$      & $\Xi_c\bar{D}^*$ & $0\frac{1}{2}^{-}$ & $18.83$ & $2.4\times 10^{-3}~(16.6\times 10^{-3})$     \\[3pt]
\hline\hline
\end{tabular*}
\end{table}

\section{Summary}\label{Sec:summary}

In this work, based on the hadronic molecular picture, we have estimated the semi-inclusive electroproduction rates of the typical multiquark states at EicC, EIC, and the proposed 24~GeV upgrade of CEBAF. 

We have employed the MC event generator Pythia to simulate the production of charmed hadron pairs at short distances. Then, the hadron pairs are bound together to form hadronic molecules through FSI at long distances. The production rates for typical tetraquarks, hidden-charm pentaquarks, and hidden-charm baryon-antibaryon molecules were estimated with such a production mechanism at EicC, EIC, and CEBAF (24~GeV). Since the c.m. energy CEBAF (24~GeV) is below the applicable energy range of Pythia, we have calculated the production rates with the electron energy at $E_e=60$, 80, 100, and 150~GeV and then extrapolated the results to 24~GeV.

Our order-of-magnitude estimates indicate that many $T_{cc}$ states, hidden-charm pentaquarks, and hidden-charm baryon-antibaryon states can be produced at EIC. While EicC will have little chance to observe double-charm $T_{cc}$ states, lots of events can be collected in the study of hidden-charm  pentaquarks and certain baryon-antibaryon molecules such as the $\Lambda_c\bar{\Lambda}_c$ molecular state. In addition, the high luminosity of the proposed 24~GeV upgrade allows CEBAF to play an important role in the search for the hidden-charm tetraquarks and pentaquarks.

\begin{acknowledgments}
We are grateful to Jorgivan Morais Dias, Shu-Ming Wu, Mao-Jun Yan, and Zhen-Hua Zhang for useful discussions.
The numerical calculations were done at the HPC Cluster of ITP-CAS.
This work is supported in part by the Chinese Academy of Sciences under Grant 
No.~XDB34030000; by the National Natural Science Foundation of China (NSFC) under 
Grants No.~12125507,  No.~11835015, No.~12047503, No.~11961141012, and No.~12275046; and by the NSFC and the Deutsche Forschungsgemeinschaft 
(DFG) through the funds provided to the TRR110 ``Symmetries and the Emergence of Structure in QCD'' (NSFC Grant No. 12070131001, DFG Project-ID 196253076); and by the Natural Science Foundation of Sichuan Province under Grant No. 2022NSFSC1795.
\end{acknowledgments}

\appendix

\section{cross section formula for the spin-3/2 molecule composed of a baryon and vector meson} \label{Sec:BV_cross_section}

The differential cross section for the inclusive production of a $BV$ pair in the c.m. frame, where $B$ and $V$ denote a baryon with spin $S=1/2$ and a vector meson, respectively, is
\begin{align}
    &d\sigma[\all + BV^*(k)]_{\text{MC}}\nonumber\\[3pt]
    =&\intMC d\phi_{\all+BV}\text{Tr}\left({\cal M}_i[\all+BV]\bar{u}[B]u[B]{\cal M}_j^*[\all+BV]\right)\epsilon^i[V]\epsilon^{*j}[V]\pairMk\nonumber\\[3pt]
    \simeq&2m_{B}\intMC d\phi_{\all+BV}\text{Tr}\left({\cal M}^i[\all+BV]{\cal M}_i^*[\all+BV]\right)\pairMk,
\end{align}
where $\epsilon^i[V]$ is the polarization vector of the vector meson $V$, and $u[B]$ is the Dirac spinor of the baryon $B$. $m_{B}$, $m_V$, and $\mu$ are the masses of $B$ and $V$ and the reduced mass of the $BV$ pair, respectively.
The cross section for the inclusive production of a $BV$ molecule with spin $S=3/2$, denoted by $X_1$, is
\begin{align}
    &\sigma[X_1+\all]\nonumber\\[3pt]
    =&\intMC d\phi_{\all+BV}\Tr \left\{{\cal M}_i[\all+BV](\slashed{k}_1+m_{B})\delta^{ii'}2m_{B}G_{\text R}g_X\bar{u}_{i'}[X_1]\right.\nonumber\\[3pt]
    &\left.u_{j'}[X_1]g_X\delta^{jj'}(\slashed{k}_1+m_{B})G_{\text R}^*{\cal M}_j^*[\all+BV]\right\}\pairMk\nonumber\\[3pt]
    \simeq&\frac{2m_{X_1}4m_{B}^2}{16m_{V}^2m_{B}^2}\intMC d\phi_{\all+BV}\Tr \left[{\cal M}_j[\all+BV]P^{(3/2)}{}^{ij}{\cal M}_i^*[\all+BV]\right]\left|G(E,\Lambda)g_X\right|^2\pairMk\nonumber\\[3pt]
    =&\frac{2}{3}\frac{m_{X_1}}{2m_{V}^2}\intMC d\phi_{\all+BV}\Tr \left[{\cal M}^i[\all+BV]{\cal M}_i^*[\all+BV]\right]\left|G(E,\Lambda)g_X\right|^2\pairMk\nonumber\\[3pt]
    \simeq&\frac{m_{X_1}}{3m_{V}m_{B}}\left|G(E,\Lambda)g_{\text{eff}}\right|^2\left(\frac{d\sigma[\all+BV]}{dk}\right)_{\text{MC}}\frac{4\pi^2\mu}{k^2},\label{Eq:cross_PV_3}
\end{align}
where $G_{\text R}$ is the relativistic scalar two-point loop function, $m_X$ is the mass of $X_1$, $k_1$ is the four-momentum of the baryon $B$, $\geff$ is the effective coupling constant in Eq.~\eqref{Eq:coupling}, and $G(E,\Lambda)$ is the Green's function in Eq.~\eqref{Eq:green_function}. $u_i[X_1]$ is the spinor of $X_1$, and $P^{(3/2)}_{ij}$ is the nonrelativistic projection operator of a spin-3/2 particle~\cite{Chung:1971ri}, 
\begin{align}
    P^{(3/2)}_{ij}=\frac{2}{3}\delta_{ij}-\frac{i}{3}\varepsilon_{ijk}\sigma^k.
\end{align}
Matching Eq.~\eqref{Eq:cross_PV_3} with the general cross section formula of Eq.~\eqref{Eq:cross_section_ep}, one can get ${\cal N}=2/3$ for the production of $X_1$. Using the same method, one can get ${\cal N}=1$ for the other hadronic molecules discussed in this work.

\bibliography{EIC_ref}

\begin{thebibliography}{84}%
\makeatletter
\providecommand \@ifxundefined [1]{%
 \@ifx{#1\undefined}
}%
\providecommand \@ifnum [1]{%
 \ifnum #1\expandafter \@firstoftwo
 \else \expandafter \@secondoftwo
 \fi
}%
\providecommand \@ifx [1]{%
 \ifx #1\expandafter \@firstoftwo
 \else \expandafter \@secondoftwo
 \fi
}%
\providecommand \natexlab [1]{#1}%
\providecommand \enquote  [1]{``#1''}%
\providecommand \bibnamefont  [1]{#1}%
\providecommand \bibfnamefont [1]{#1}%
\providecommand \citenamefont [1]{#1}%
\providecommand \href@noop [0]{\@secondoftwo}%
\providecommand \href [0]{\begingroup \@sanitize@url \@href}%
\providecommand \@href[1]{\@@startlink{#1}\@@href}%
\providecommand \@@href[1]{\endgroup#1\@@endlink}%
\providecommand \@sanitize@url [0]{\catcode `\\12\catcode `\$12\catcode
  `\&12\catcode `\#12\catcode `\^12\catcode `\_12\catcode `\%12\relax}%
\providecommand \@@startlink[1]{}%
\providecommand \@@endlink[0]{}%
\providecommand \url  [0]{\begingroup\@sanitize@url \@url }%
\providecommand \@url [1]{\endgroup\@href {#1}{\urlprefix }}%
\providecommand \urlprefix  [0]{URL }%
\providecommand \Eprint [0]{\href }%
\providecommand \doibase [0]{http://dx.doi.org/}%
\providecommand \selectlanguage [0]{\@gobble}%
\providecommand \bibinfo  [0]{\@secondoftwo}%
\providecommand \bibfield  [0]{\@secondoftwo}%
\providecommand \translation [1]{[#1]}%
\providecommand \BibitemOpen [0]{}%
\providecommand \bibitemStop [0]{}%
\providecommand \bibitemNoStop [0]{.\EOS\space}%
\providecommand \EOS [0]{\spacefactor3000\relax}%
\providecommand \BibitemShut  [1]{\csname bibitem#1\endcsname}%
\let\auto@bib@innerbib\@empty
\bibitem [{\citenamefont {Ablikim}\ \emph {et~al.}(2020)\citenamefont {Ablikim}
  \emph {et~al.}}]{BESIII:2020nme}%
  \BibitemOpen
  \bibfield  {author} {\bibinfo {author} {\bibfnamefont {M.}~\bibnamefont
  {Ablikim}} \emph {et~al.} (\bibinfo {collaboration} {BESIII}),\ }\href
  {\doibase 10.1088/1674-1137/44/4/040001} {\bibfield  {journal} {\bibinfo
  {journal} {Chin. Phys. C}\ }\textbf {\bibinfo {volume} {44}},\ \bibinfo
  {pages} {040001} (\bibinfo {year} {2020})},\ \Eprint
  {http://arxiv.org/abs/1912.05983} {arXiv:1912.05983 [hep-ex]} \BibitemShut
  {NoStop}%
\bibitem [{\citenamefont {Cerri}\ \emph {et~al.}(2019)\citenamefont {Cerri}
  \emph {et~al.}}]{Cerri:2018ypt}%
  \BibitemOpen
  \bibfield  {author} {\bibinfo {author} {\bibfnamefont {A.}~\bibnamefont
  {Cerri}} \emph {et~al.},\ }\href {\doibase 10.23731/CYRM-2019-007.867}
  {\bibfield  {journal} {\bibinfo  {journal} {CERN Yellow Rep. Monogr.}\
  }\textbf {\bibinfo {volume} {7}},\ \bibinfo {pages} {867} (\bibinfo {year}
  {2019})},\ \Eprint {http://arxiv.org/abs/1812.07638} {arXiv:1812.07638
  [hep-ph]} \BibitemShut {NoStop}%
\bibitem [{\citenamefont {Altmannshofer}\ \emph {et~al.}(2019)\citenamefont
  {Altmannshofer} \emph {et~al.}}]{Belle-II:2018jsg}%
  \BibitemOpen
  \bibfield  {author} {\bibinfo {author} {\bibfnamefont {W.}~\bibnamefont
  {Altmannshofer}} \emph {et~al.} (\bibinfo {collaboration} {Belle-II}),\
  }\href {\doibase 10.1093/ptep/ptz106} {\bibfield  {journal} {\bibinfo
  {journal} {PTEP}\ }\textbf {\bibinfo {volume} {2019}},\ \bibinfo {pages}
  {123C01} (\bibinfo {year} {2019})},\ \bibinfo {note} {[Erratum: PTEP 2020,
  029201 (2020)]},\ \Eprint {http://arxiv.org/abs/1808.10567} {arXiv:1808.10567
  [hep-ex]} \BibitemShut {NoStop}%
\bibitem [{\citenamefont {Chen}\ \emph {et~al.}(2016)\citenamefont {Chen},
  \citenamefont {Chen}, \citenamefont {Liu},\ and\ \citenamefont
  {Zhu}}]{Chen:2016qju}%
  \BibitemOpen
  \bibfield  {author} {\bibinfo {author} {\bibfnamefont {H.-X.}\ \bibnamefont
  {Chen}}, \bibinfo {author} {\bibfnamefont {W.}~\bibnamefont {Chen}}, \bibinfo
  {author} {\bibfnamefont {X.}~\bibnamefont {Liu}}, \ and\ \bibinfo {author}
  {\bibfnamefont {S.-L.}\ \bibnamefont {Zhu}},\ }\href {\doibase
  10.1016/j.physrep.2016.05.004} {\bibfield  {journal} {\bibinfo  {journal}
  {Phys. Rep.}\ }\textbf {\bibinfo {volume} {639}},\ \bibinfo {pages} {1}
  (\bibinfo {year} {2016})},\ \Eprint {http://arxiv.org/abs/1601.02092}
  {arXiv:1601.02092 [hep-ph]} \BibitemShut {NoStop}%
\bibitem [{\citenamefont {Hosaka}\ \emph {et~al.}(2016)\citenamefont {Hosaka},
  \citenamefont {Iijima}, \citenamefont {Miyabayashi}, \citenamefont {Sakai},\
  and\ \citenamefont {Yasui}}]{Hosaka:2016pey}%
  \BibitemOpen
  \bibfield  {author} {\bibinfo {author} {\bibfnamefont {A.}~\bibnamefont
  {Hosaka}}, \bibinfo {author} {\bibfnamefont {T.}~\bibnamefont {Iijima}},
  \bibinfo {author} {\bibfnamefont {K.}~\bibnamefont {Miyabayashi}}, \bibinfo
  {author} {\bibfnamefont {Y.}~\bibnamefont {Sakai}}, \ and\ \bibinfo {author}
  {\bibfnamefont {S.}~\bibnamefont {Yasui}},\ }\href {\doibase
  10.1093/ptep/ptw045} {\bibfield  {journal} {\bibinfo  {journal} {Prog. Theor.
  Exp. Phys.}\ }\textbf {\bibinfo {volume} {2016}},\ \bibinfo {pages} {062C01}
  (\bibinfo {year} {2016})},\ \Eprint {http://arxiv.org/abs/1603.09229}
  {arXiv:1603.09229 [hep-ph]} \BibitemShut {NoStop}%
\bibitem [{\citenamefont {Lebed}\ \emph {et~al.}(2017)\citenamefont {Lebed},
  \citenamefont {Mitchell},\ and\ \citenamefont {Swanson}}]{Lebed:2016hpi}%
  \BibitemOpen
  \bibfield  {author} {\bibinfo {author} {\bibfnamefont {R.~F.}\ \bibnamefont
  {Lebed}}, \bibinfo {author} {\bibfnamefont {R.~E.}\ \bibnamefont {Mitchell}},
  \ and\ \bibinfo {author} {\bibfnamefont {E.~S.}\ \bibnamefont {Swanson}},\
  }\href {\doibase 10.1016/j.ppnp.2016.11.003} {\bibfield  {journal} {\bibinfo
  {journal} {Prog. Part. Nucl. Phys.}\ }\textbf {\bibinfo {volume} {93}},\
  \bibinfo {pages} {143} (\bibinfo {year} {2017})},\ \Eprint
  {http://arxiv.org/abs/1610.04528} {arXiv:1610.04528 [hep-ph]} \BibitemShut
  {NoStop}%
\bibitem [{\citenamefont {Esposito}\ \emph {et~al.}(2016)\citenamefont
  {Esposito}, \citenamefont {Pilloni},\ and\ \citenamefont
  {Polosa}}]{Esposito:2016noz}%
  \BibitemOpen
  \bibfield  {author} {\bibinfo {author} {\bibfnamefont {A.}~\bibnamefont
  {Esposito}}, \bibinfo {author} {\bibfnamefont {A.}~\bibnamefont {Pilloni}}, \
  and\ \bibinfo {author} {\bibfnamefont {A.~D.}\ \bibnamefont {Polosa}},\
  }\href {\doibase 10.1016/j.physrep.2016.11.002} {\bibfield  {journal}
  {\bibinfo  {journal} {Phys. Rep.}\ }\textbf {\bibinfo {volume} {668}},\
  \bibinfo {pages} {1} (\bibinfo {year} {2016})},\ \Eprint
  {http://arxiv.org/abs/1611.07920} {arXiv:1611.07920 [hep-ph]} \BibitemShut
  {NoStop}%
\bibitem [{\citenamefont {Guo}\ \emph {et~al.}(2018)\citenamefont {Guo},
  \citenamefont {Hanhart}, \citenamefont {Mei{\ss}ner}, \citenamefont {Wang},
  \citenamefont {Zhao},\ and\ \citenamefont {Zou}}]{Guo:2017jvc}%
  \BibitemOpen
  \bibfield  {author} {\bibinfo {author} {\bibfnamefont {F.-K.}\ \bibnamefont
  {Guo}}, \bibinfo {author} {\bibfnamefont {C.}~\bibnamefont {Hanhart}},
  \bibinfo {author} {\bibfnamefont {U.-G.}\ \bibnamefont {Mei{\ss}ner}},
  \bibinfo {author} {\bibfnamefont {Q.}~\bibnamefont {Wang}}, \bibinfo {author}
  {\bibfnamefont {Q.}~\bibnamefont {Zhao}}, \ and\ \bibinfo {author}
  {\bibfnamefont {B.-S.}\ \bibnamefont {Zou}},\ }\href {\doibase
  10.1103/RevModPhys.90.015004} {\bibfield  {journal} {\bibinfo  {journal}
  {Rev. Mod. Phys.}\ }\textbf {\bibinfo {volume} {90}},\ \bibinfo {pages}
  {015004} (\bibinfo {year} {2018})},\ \Eprint
  {http://arxiv.org/abs/1705.00141} {arXiv:1705.00141 [hep-ph]} \BibitemShut
  {NoStop}%
\bibitem [{\citenamefont {Olsen}\ \emph {et~al.}(2018)\citenamefont {Olsen},
  \citenamefont {Skwarnicki},\ and\ \citenamefont {Zieminska}}]{Olsen:2017bmm}%
  \BibitemOpen
  \bibfield  {author} {\bibinfo {author} {\bibfnamefont {S.~L.}\ \bibnamefont
  {Olsen}}, \bibinfo {author} {\bibfnamefont {T.}~\bibnamefont {Skwarnicki}}, \
  and\ \bibinfo {author} {\bibfnamefont {D.}~\bibnamefont {Zieminska}},\ }\href
  {\doibase 10.1103/RevModPhys.90.015003} {\bibfield  {journal} {\bibinfo
  {journal} {Rev. Mod. Phys.}\ }\textbf {\bibinfo {volume} {90}},\ \bibinfo
  {pages} {015003} (\bibinfo {year} {2018})},\ \Eprint
  {http://arxiv.org/abs/1708.04012} {arXiv:1708.04012 [hep-ph]} \BibitemShut
  {NoStop}%
\bibitem [{\citenamefont {Liu}\ \emph {et~al.}(2019{\natexlab{a}})\citenamefont
  {Liu}, \citenamefont {Chen}, \citenamefont {Chen}, \citenamefont {Liu},\ and\
  \citenamefont {Zhu}}]{Liu:2019zoy}%
  \BibitemOpen
  \bibfield  {author} {\bibinfo {author} {\bibfnamefont {Y.-R.}\ \bibnamefont
  {Liu}}, \bibinfo {author} {\bibfnamefont {H.-X.}\ \bibnamefont {Chen}},
  \bibinfo {author} {\bibfnamefont {W.}~\bibnamefont {Chen}}, \bibinfo {author}
  {\bibfnamefont {X.}~\bibnamefont {Liu}}, \ and\ \bibinfo {author}
  {\bibfnamefont {S.-L.}\ \bibnamefont {Zhu}},\ }\href {\doibase
  10.1016/j.ppnp.2019.04.003} {\bibfield  {journal} {\bibinfo  {journal} {Prog.
  Part. Nucl. Phys.}\ }\textbf {\bibinfo {volume} {107}},\ \bibinfo {pages}
  {237} (\bibinfo {year} {2019}{\natexlab{a}})},\ \Eprint
  {http://arxiv.org/abs/1903.11976} {arXiv:1903.11976 [hep-ph]} \BibitemShut
  {NoStop}%
\bibitem [{\citenamefont {Brambilla}\ \emph {et~al.}(2020)\citenamefont
  {Brambilla}, \citenamefont {Eidelman}, \citenamefont {Hanhart}, \citenamefont
  {Nefediev}, \citenamefont {Shen}, \citenamefont {Thomas}, \citenamefont
  {Vairo},\ and\ \citenamefont {Yuan}}]{Brambilla:2019esw}%
  \BibitemOpen
  \bibfield  {author} {\bibinfo {author} {\bibfnamefont {N.}~\bibnamefont
  {Brambilla}}, \bibinfo {author} {\bibfnamefont {S.}~\bibnamefont {Eidelman}},
  \bibinfo {author} {\bibfnamefont {C.}~\bibnamefont {Hanhart}}, \bibinfo
  {author} {\bibfnamefont {A.}~\bibnamefont {Nefediev}}, \bibinfo {author}
  {\bibfnamefont {C.-P.}\ \bibnamefont {Shen}}, \bibinfo {author}
  {\bibfnamefont {C.~E.}\ \bibnamefont {Thomas}}, \bibinfo {author}
  {\bibfnamefont {A.}~\bibnamefont {Vairo}}, \ and\ \bibinfo {author}
  {\bibfnamefont {C.-Z.}\ \bibnamefont {Yuan}},\ }\href {\doibase
  10.1016/j.physrep.2020.05.001} {\bibfield  {journal} {\bibinfo  {journal}
  {Phys. Rep.}\ }\textbf {\bibinfo {volume} {873}},\ \bibinfo {pages} {1}
  (\bibinfo {year} {2020})},\ \Eprint {http://arxiv.org/abs/1907.07583}
  {arXiv:1907.07583 [hep-ex]} \BibitemShut {NoStop}%
\bibitem [{\citenamefont {Guo}\ \emph {et~al.}(2020)\citenamefont {Guo},
  \citenamefont {Liu},\ and\ \citenamefont {Sakai}}]{Guo:2019twa}%
  \BibitemOpen
  \bibfield  {author} {\bibinfo {author} {\bibfnamefont {F.-K.}\ \bibnamefont
  {Guo}}, \bibinfo {author} {\bibfnamefont {X.-H.}\ \bibnamefont {Liu}}, \ and\
  \bibinfo {author} {\bibfnamefont {S.}~\bibnamefont {Sakai}},\ }\href
  {\doibase 10.1016/j.ppnp.2020.103757} {\bibfield  {journal} {\bibinfo
  {journal} {Prog. Part. Nucl. Phys.}\ }\textbf {\bibinfo {volume} {112}},\
  \bibinfo {pages} {103757} (\bibinfo {year} {2020})},\ \Eprint
  {http://arxiv.org/abs/1912.07030} {arXiv:1912.07030 [hep-ph]} \BibitemShut
  {NoStop}%
\bibitem [{\citenamefont {Chen}\ \emph {et~al.}(2022)\citenamefont {Chen},
  \citenamefont {Chen}, \citenamefont {Liu}, \citenamefont {Liu},\ and\
  \citenamefont {Zhu}}]{Chen:2022asf}%
  \BibitemOpen
  \bibfield  {author} {\bibinfo {author} {\bibfnamefont {H.-X.}\ \bibnamefont
  {Chen}}, \bibinfo {author} {\bibfnamefont {W.}~\bibnamefont {Chen}}, \bibinfo
  {author} {\bibfnamefont {X.}~\bibnamefont {Liu}}, \bibinfo {author}
  {\bibfnamefont {Y.-R.}\ \bibnamefont {Liu}}, \ and\ \bibinfo {author}
  {\bibfnamefont {S.-L.}\ \bibnamefont {Zhu}},\ }\href@noop {} {\  (\bibinfo
  {year} {2022})},\ \Eprint {http://arxiv.org/abs/2204.02649} {arXiv:2204.02649
  [hep-ph]} \BibitemShut {NoStop}%
\bibitem [{\citenamefont {Adolph}\ \emph {et~al.}(2015)\citenamefont {Adolph}
  \emph {et~al.}}]{COMPASS:2014mhq}%
  \BibitemOpen
  \bibfield  {author} {\bibinfo {author} {\bibfnamefont {C.}~\bibnamefont
  {Adolph}} \emph {et~al.} (\bibinfo {collaboration} {COMPASS}),\ }\href
  {\doibase 10.1016/j.physletb.2015.01.042} {\bibfield  {journal} {\bibinfo
  {journal} {Phys. Lett. B}\ }\textbf {\bibinfo {volume} {742}},\ \bibinfo
  {pages} {330} (\bibinfo {year} {2015})},\ \Eprint
  {http://arxiv.org/abs/1407.6186} {arXiv:1407.6186 [hep-ex]} \BibitemShut
  {NoStop}%
\bibitem [{\citenamefont {Aghasyan}\ \emph {et~al.}(2018)\citenamefont
  {Aghasyan} \emph {et~al.}}]{COMPASS:2017wql}%
  \BibitemOpen
  \bibfield  {author} {\bibinfo {author} {\bibfnamefont {M.}~\bibnamefont
  {Aghasyan}} \emph {et~al.} (\bibinfo {collaboration} {COMPASS}),\ }\href
  {\doibase 10.1016/j.physletb.2018.07.008} {\bibfield  {journal} {\bibinfo
  {journal} {Phys. Lett. B}\ }\textbf {\bibinfo {volume} {783}},\ \bibinfo
  {pages} {334} (\bibinfo {year} {2018})},\ \Eprint
  {http://arxiv.org/abs/1707.01796} {arXiv:1707.01796 [hep-ex]} \BibitemShut
  {NoStop}%
\bibitem [{\citenamefont {Ali}\ \emph {et~al.}(2019)\citenamefont {Ali} \emph
  {et~al.}}]{GlueX:2019mkq}%
  \BibitemOpen
  \bibfield  {author} {\bibinfo {author} {\bibfnamefont {A.}~\bibnamefont
  {Ali}} \emph {et~al.} (\bibinfo {collaboration} {GlueX}),\ }\href {\doibase
  10.1103/PhysRevLett.123.072001} {\bibfield  {journal} {\bibinfo  {journal}
  {Phys. Rev. Lett.}\ }\textbf {\bibinfo {volume} {123}},\ \bibinfo {pages}
  {072001} (\bibinfo {year} {2019})},\ \Eprint
  {http://arxiv.org/abs/1905.10811} {arXiv:1905.10811 [nucl-ex]} \BibitemShut
  {NoStop}%
\bibitem [{\citenamefont {Anderle}\ \emph {et~al.}(2021)\citenamefont {Anderle}
  \emph {et~al.}}]{Anderle:2021wcy}%
  \BibitemOpen
  \bibfield  {author} {\bibinfo {author} {\bibfnamefont {D.~P.}\ \bibnamefont
  {Anderle}} \emph {et~al.},\ }\href {\doibase 10.1007/s11467-021-1062-0}
  {\bibfield  {journal} {\bibinfo  {journal} {Front. Phys. (Beijing)}\ }\textbf
  {\bibinfo {volume} {16}},\ \bibinfo {pages} {64701} (\bibinfo {year}
  {2021})},\ \Eprint {http://arxiv.org/abs/2102.09222} {arXiv:2102.09222
  [nucl-ex]} \BibitemShut {NoStop}%
\bibitem [{\citenamefont {Abdul~Khalek}\ \emph {et~al.}(2021)\citenamefont
  {Abdul~Khalek} \emph {et~al.}}]{AbdulKhalek:2021gbh}%
  \BibitemOpen
  \bibfield  {author} {\bibinfo {author} {\bibfnamefont {R.}~\bibnamefont
  {Abdul~Khalek}} \emph {et~al.},\ }\href@noop {} {\  (\bibinfo {year}
  {2021})},\ \Eprint {http://arxiv.org/abs/2103.05419} {arXiv:2103.05419
  [physics.ins-det]} \BibitemShut {NoStop}%
\bibitem [{\citenamefont {Arrington}\ \emph {et~al.}(2021)\citenamefont
  {Arrington} \emph {et~al.}}]{Arrington:2021alx}%
  \BibitemOpen
  \bibfield  {author} {\bibinfo {author} {\bibfnamefont {J.}~\bibnamefont
  {Arrington}} \emph {et~al.},\ }\href@noop {} {\  (\bibinfo {year} {2021})},\
  \Eprint {http://arxiv.org/abs/2112.00060} {arXiv:2112.00060 [nucl-ex]}
  \BibitemShut {NoStop}%
\bibitem [{\citenamefont {Huang}\ \emph {et~al.}(2014)\citenamefont {Huang},
  \citenamefont {He}, \citenamefont {Zhang},\ and\ \citenamefont
  {Chen}}]{Huang:2013mua}%
  \BibitemOpen
  \bibfield  {author} {\bibinfo {author} {\bibfnamefont {Y.}~\bibnamefont
  {Huang}}, \bibinfo {author} {\bibfnamefont {J.}~\bibnamefont {He}}, \bibinfo
  {author} {\bibfnamefont {H.-F.}\ \bibnamefont {Zhang}}, \ and\ \bibinfo
  {author} {\bibfnamefont {X.-R.}\ \bibnamefont {Chen}},\ }\href {\doibase
  10.1088/0954-3899/41/11/115004} {\bibfield  {journal} {\bibinfo  {journal}
  {J. Phys. G}\ }\textbf {\bibinfo {volume} {41}},\ \bibinfo {pages} {115004}
  (\bibinfo {year} {2014})},\ \Eprint {http://arxiv.org/abs/1305.4434}
  {arXiv:1305.4434 [nucl-th]} \BibitemShut {NoStop}%
\bibitem [{\citenamefont {Wang}\ \emph
  {et~al.}(2015{\natexlab{a}})\citenamefont {Wang}, \citenamefont {Liu},\ and\
  \citenamefont {Zhao}}]{Wang:2015jsa}%
  \BibitemOpen
  \bibfield  {author} {\bibinfo {author} {\bibfnamefont {Q.}~\bibnamefont
  {Wang}}, \bibinfo {author} {\bibfnamefont {X.-H.}\ \bibnamefont {Liu}}, \
  and\ \bibinfo {author} {\bibfnamefont {Q.}~\bibnamefont {Zhao}},\ }\href
  {\doibase 10.1103/PhysRevD.92.034022} {\bibfield  {journal} {\bibinfo
  {journal} {Phys. Rev. D}\ }\textbf {\bibinfo {volume} {92}},\ \bibinfo
  {pages} {034022} (\bibinfo {year} {2015}{\natexlab{a}})},\ \Eprint
  {http://arxiv.org/abs/1508.00339} {arXiv:1508.00339 [hep-ph]} \BibitemShut
  {NoStop}%
\bibitem [{\citenamefont {Huang}\ \emph {et~al.}(2016)\citenamefont {Huang},
  \citenamefont {Xie}, \citenamefont {He}, \citenamefont {Chen},\ and\
  \citenamefont {Zhang}}]{Huang:2016tcr}%
  \BibitemOpen
  \bibfield  {author} {\bibinfo {author} {\bibfnamefont {Y.}~\bibnamefont
  {Huang}}, \bibinfo {author} {\bibfnamefont {J.-J.}\ \bibnamefont {Xie}},
  \bibinfo {author} {\bibfnamefont {J.}~\bibnamefont {He}}, \bibinfo {author}
  {\bibfnamefont {X.}~\bibnamefont {Chen}}, \ and\ \bibinfo {author}
  {\bibfnamefont {H.-F.}\ \bibnamefont {Zhang}},\ }\href {\doibase
  10.1088/1674-1137/40/12/124104} {\bibfield  {journal} {\bibinfo  {journal}
  {Chin. Phys. C}\ }\textbf {\bibinfo {volume} {40}},\ \bibinfo {pages}
  {124104} (\bibinfo {year} {2016})},\ \Eprint
  {http://arxiv.org/abs/1604.05969} {arXiv:1604.05969 [nucl-th]} \BibitemShut
  {NoStop}%
\bibitem [{\citenamefont {Hiller~Blin}\ \emph {et~al.}(2016)\citenamefont
  {Hiller~Blin}, \citenamefont {Fern\'andez-Ram\'\i{}rez}, \citenamefont
  {Jackura}, \citenamefont {Mathieu}, \citenamefont {Mokeev}, \citenamefont
  {Pilloni},\ and\ \citenamefont {Szczepaniak}}]{HillerBlin:2016odx}%
  \BibitemOpen
  \bibfield  {author} {\bibinfo {author} {\bibfnamefont {A.~N.}\ \bibnamefont
  {Hiller~Blin}}, \bibinfo {author} {\bibfnamefont {C.}~\bibnamefont
  {Fern\'andez-Ram\'\i{}rez}}, \bibinfo {author} {\bibfnamefont
  {A.}~\bibnamefont {Jackura}}, \bibinfo {author} {\bibfnamefont
  {V.}~\bibnamefont {Mathieu}}, \bibinfo {author} {\bibfnamefont {V.~I.}\
  \bibnamefont {Mokeev}}, \bibinfo {author} {\bibfnamefont {A.}~\bibnamefont
  {Pilloni}}, \ and\ \bibinfo {author} {\bibfnamefont {A.~P.}\ \bibnamefont
  {Szczepaniak}},\ }\href {\doibase 10.1103/PhysRevD.94.034002} {\bibfield
  {journal} {\bibinfo  {journal} {Phys. Rev. D}\ }\textbf {\bibinfo {volume}
  {94}},\ \bibinfo {pages} {034002} (\bibinfo {year} {2016})},\ \Eprint
  {http://arxiv.org/abs/1606.08912} {arXiv:1606.08912 [hep-ph]} \BibitemShut
  {NoStop}%
\bibitem [{\citenamefont {Karliner}\ and\ \citenamefont
  {Rosner}(2016)}]{Karliner:2015voa}%
  \BibitemOpen
  \bibfield  {author} {\bibinfo {author} {\bibfnamefont {M.}~\bibnamefont
  {Karliner}}\ and\ \bibinfo {author} {\bibfnamefont {J.~L.}\ \bibnamefont
  {Rosner}},\ }\href {\doibase 10.1016/j.physletb.2015.11.068} {\bibfield
  {journal} {\bibinfo  {journal} {Phys. Lett. B}\ }\textbf {\bibinfo {volume}
  {752}},\ \bibinfo {pages} {329} (\bibinfo {year} {2016})},\ \Eprint
  {http://arxiv.org/abs/1508.01496} {arXiv:1508.01496 [hep-ph]} \BibitemShut
  {NoStop}%
\bibitem [{\citenamefont {Kubarovsky}\ and\ \citenamefont
  {Voloshin}(2015)}]{Kubarovsky:2015aaa}%
  \BibitemOpen
  \bibfield  {author} {\bibinfo {author} {\bibfnamefont {V.}~\bibnamefont
  {Kubarovsky}}\ and\ \bibinfo {author} {\bibfnamefont {M.~B.}\ \bibnamefont
  {Voloshin}},\ }\href {\doibase 10.1103/PhysRevD.92.031502} {\bibfield
  {journal} {\bibinfo  {journal} {Phys. Rev. D}\ }\textbf {\bibinfo {volume}
  {92}},\ \bibinfo {pages} {031502} (\bibinfo {year} {2015})},\ \Eprint
  {http://arxiv.org/abs/1508.00888} {arXiv:1508.00888 [hep-ph]} \BibitemShut
  {NoStop}%
\bibitem [{\citenamefont {Winney}\ \emph {et~al.}(2019)\citenamefont {Winney},
  \citenamefont {Fanelli}, \citenamefont {Pilloni}, \citenamefont
  {Hiller~Blin}, \citenamefont {Fern\'andez-Ram\'\i{}rez}, \citenamefont
  {Albaladejo}, \citenamefont {Mathieu}, \citenamefont {Mokeev},\ and\
  \citenamefont {Szczepaniak}}]{Winney:2019edt}%
  \BibitemOpen
  \bibfield  {author} {\bibinfo {author} {\bibfnamefont {D.}~\bibnamefont
  {Winney}}, \bibinfo {author} {\bibfnamefont {C.}~\bibnamefont {Fanelli}},
  \bibinfo {author} {\bibfnamefont {A.}~\bibnamefont {Pilloni}}, \bibinfo
  {author} {\bibfnamefont {A.~N.}\ \bibnamefont {Hiller~Blin}}, \bibinfo
  {author} {\bibfnamefont {C.}~\bibnamefont {Fern\'andez-Ram\'\i{}rez}},
  \bibinfo {author} {\bibfnamefont {M.}~\bibnamefont {Albaladejo}}, \bibinfo
  {author} {\bibfnamefont {V.}~\bibnamefont {Mathieu}}, \bibinfo {author}
  {\bibfnamefont {V.~I.}\ \bibnamefont {Mokeev}}, \ and\ \bibinfo {author}
  {\bibfnamefont {A.~P.}\ \bibnamefont {Szczepaniak}} (\bibinfo {collaboration}
  {JPAC}),\ }\href {\doibase 10.1103/PhysRevD.100.034019} {\bibfield  {journal}
  {\bibinfo  {journal} {Phys. Rev. D}\ }\textbf {\bibinfo {volume} {100}},\
  \bibinfo {pages} {034019} (\bibinfo {year} {2019})},\ \Eprint
  {http://arxiv.org/abs/1907.09393} {arXiv:1907.09393 [hep-ph]} \BibitemShut
  {NoStop}%
\bibitem [{\citenamefont {Paryev}\ and\ \citenamefont
  {Kiselev}(2018)}]{Paryev:2018fyv}%
  \BibitemOpen
  \bibfield  {author} {\bibinfo {author} {\bibfnamefont {E.~Y.}\ \bibnamefont
  {Paryev}}\ and\ \bibinfo {author} {\bibfnamefont {Y.~T.}\ \bibnamefont
  {Kiselev}},\ }\href {\doibase 10.1016/j.nuclphysa.2018.08.009} {\bibfield
  {journal} {\bibinfo  {journal} {Nucl. Phys. A}\ }\textbf {\bibinfo {volume}
  {978}},\ \bibinfo {pages} {201} (\bibinfo {year} {2018})},\ \Eprint
  {http://arxiv.org/abs/1810.01715} {arXiv:1810.01715 [nucl-th]} \BibitemShut
  {NoStop}%
\bibitem [{\citenamefont {Wang}\ \emph {et~al.}(2019)\citenamefont {Wang},
  \citenamefont {Chen},\ and\ \citenamefont {He}}]{Wang:2019krd}%
  \BibitemOpen
  \bibfield  {author} {\bibinfo {author} {\bibfnamefont {X.-Y.}\ \bibnamefont
  {Wang}}, \bibinfo {author} {\bibfnamefont {X.-R.}\ \bibnamefont {Chen}}, \
  and\ \bibinfo {author} {\bibfnamefont {J.}~\bibnamefont {He}},\ }\href
  {\doibase 10.1103/PhysRevD.99.114007} {\bibfield  {journal} {\bibinfo
  {journal} {Phys. Rev. D}\ }\textbf {\bibinfo {volume} {99}},\ \bibinfo
  {pages} {114007} (\bibinfo {year} {2019})},\ \Eprint
  {http://arxiv.org/abs/1904.11706} {arXiv:1904.11706 [hep-ph]} \BibitemShut
  {NoStop}%
\bibitem [{\citenamefont {Gon\c{c}alves}\ and\ \citenamefont
  {Jaime}(2020)}]{Goncalves:2019vvo}%
  \BibitemOpen
  \bibfield  {author} {\bibinfo {author} {\bibfnamefont {V.~P.}\ \bibnamefont
  {Gon\c{c}alves}}\ and\ \bibinfo {author} {\bibfnamefont {M.~M.}\ \bibnamefont
  {Jaime}},\ }\href {\doibase 10.1016/j.physletb.2020.135447} {\bibfield
  {journal} {\bibinfo  {journal} {Phys. Lett. B}\ }\textbf {\bibinfo {volume}
  {805}},\ \bibinfo {pages} {135447} (\bibinfo {year} {2020})},\ \Eprint
  {http://arxiv.org/abs/1911.10886} {arXiv:1911.10886 [hep-ph]} \BibitemShut
  {NoStop}%
\bibitem [{\citenamefont {Wu}\ \emph {et~al.}(2019)\citenamefont {Wu},
  \citenamefont {Lee},\ and\ \citenamefont {Zou}}]{Wu:2019adv}%
  \BibitemOpen
  \bibfield  {author} {\bibinfo {author} {\bibfnamefont {J.-J.}\ \bibnamefont
  {Wu}}, \bibinfo {author} {\bibfnamefont {T.~S.~H.}\ \bibnamefont {Lee}}, \
  and\ \bibinfo {author} {\bibfnamefont {B.-S.}\ \bibnamefont {Zou}},\ }\href
  {\doibase 10.1103/PhysRevC.100.035206} {\bibfield  {journal} {\bibinfo
  {journal} {Phys. Rev. C}\ }\textbf {\bibinfo {volume} {100}},\ \bibinfo
  {pages} {035206} (\bibinfo {year} {2019})},\ \Eprint
  {http://arxiv.org/abs/1906.05375} {arXiv:1906.05375 [nucl-th]} \BibitemShut
  {NoStop}%
\bibitem [{\citenamefont {Xie}\ \emph {et~al.}(2021)\citenamefont {Xie},
  \citenamefont {Cao}, \citenamefont {Liang},\ and\ \citenamefont
  {Chen}}]{Xie:2020niw}%
  \BibitemOpen
  \bibfield  {author} {\bibinfo {author} {\bibfnamefont {Y.-P.}\ \bibnamefont
  {Xie}}, \bibinfo {author} {\bibfnamefont {X.}~\bibnamefont {Cao}}, \bibinfo
  {author} {\bibfnamefont {Y.-T.}\ \bibnamefont {Liang}}, \ and\ \bibinfo
  {author} {\bibfnamefont {X.}~\bibnamefont {Chen}},\ }\href {\doibase
  10.1088/1674-1137/abdea9} {\bibfield  {journal} {\bibinfo  {journal} {Chin.
  Phys. C}\ }\textbf {\bibinfo {volume} {45}},\ \bibinfo {pages} {043105}
  (\bibinfo {year} {2021})},\ \Eprint {http://arxiv.org/abs/2003.11729}
  {arXiv:2003.11729 [hep-ph]} \BibitemShut {NoStop}%
\bibitem [{\citenamefont {Yang}\ \emph {et~al.}(2020)\citenamefont {Yang},
  \citenamefont {Cao}, \citenamefont {Liang},\ and\ \citenamefont
  {Wu}}]{Yang:2020eye}%
  \BibitemOpen
  \bibfield  {author} {\bibinfo {author} {\bibfnamefont {Z.}~\bibnamefont
  {Yang}}, \bibinfo {author} {\bibfnamefont {X.}~\bibnamefont {Cao}}, \bibinfo
  {author} {\bibfnamefont {Y.-T.}\ \bibnamefont {Liang}}, \ and\ \bibinfo
  {author} {\bibfnamefont {J.-J.}\ \bibnamefont {Wu}},\ }\href {\doibase
  10.1088/1674-1137/44/8/084102} {\bibfield  {journal} {\bibinfo  {journal}
  {Chin. Phys. C}\ }\textbf {\bibinfo {volume} {44}},\ \bibinfo {pages}
  {084102} (\bibinfo {year} {2020})},\ \Eprint
  {http://arxiv.org/abs/2003.06774} {arXiv:2003.06774 [hep-ph]} \BibitemShut
  {NoStop}%
\bibitem [{\citenamefont {Liu}\ \emph {et~al.}(2008)\citenamefont {Liu},
  \citenamefont {Zhao},\ and\ \citenamefont {Close}}]{Liu:2008qx}%
  \BibitemOpen
  \bibfield  {author} {\bibinfo {author} {\bibfnamefont {X.-H.}\ \bibnamefont
  {Liu}}, \bibinfo {author} {\bibfnamefont {Q.}~\bibnamefont {Zhao}}, \ and\
  \bibinfo {author} {\bibfnamefont {F.~E.}\ \bibnamefont {Close}},\ }\href
  {\doibase 10.1103/PhysRevD.77.094005} {\bibfield  {journal} {\bibinfo
  {journal} {Phys. Rev. D}\ }\textbf {\bibinfo {volume} {77}},\ \bibinfo
  {pages} {094005} (\bibinfo {year} {2008})},\ \Eprint
  {http://arxiv.org/abs/0802.2648} {arXiv:0802.2648 [hep-ph]} \BibitemShut
  {NoStop}%
\bibitem [{\citenamefont {Galata}(2011)}]{Galata:2011bi}%
  \BibitemOpen
  \bibfield  {author} {\bibinfo {author} {\bibfnamefont {G.}~\bibnamefont
  {Galata}},\ }\href {\doibase 10.1103/PhysRevC.83.065203} {\bibfield
  {journal} {\bibinfo  {journal} {Phys. Rev. C}\ }\textbf {\bibinfo {volume}
  {83}},\ \bibinfo {pages} {065203} (\bibinfo {year} {2011})},\ \Eprint
  {http://arxiv.org/abs/1102.2070} {arXiv:1102.2070 [hep-ph]} \BibitemShut
  {NoStop}%
\bibitem [{\citenamefont {Lin}\ \emph {et~al.}(2013)\citenamefont {Lin},
  \citenamefont {Liu},\ and\ \citenamefont {Xu}}]{Lin:2013mka}%
  \BibitemOpen
  \bibfield  {author} {\bibinfo {author} {\bibfnamefont {Q.-Y.}\ \bibnamefont
  {Lin}}, \bibinfo {author} {\bibfnamefont {X.}~\bibnamefont {Liu}}, \ and\
  \bibinfo {author} {\bibfnamefont {H.-S.}\ \bibnamefont {Xu}},\ }\href
  {\doibase 10.1103/PhysRevD.88.114009} {\bibfield  {journal} {\bibinfo
  {journal} {Phys. Rev. D}\ }\textbf {\bibinfo {volume} {88}},\ \bibinfo
  {pages} {114009} (\bibinfo {year} {2013})},\ \Eprint
  {http://arxiv.org/abs/1308.6345} {arXiv:1308.6345 [hep-ph]} \BibitemShut
  {NoStop}%
\bibitem [{\citenamefont {Lin}\ \emph {et~al.}(2014)\citenamefont {Lin},
  \citenamefont {Liu},\ and\ \citenamefont {Xu}}]{Lin:2013ppa}%
  \BibitemOpen
  \bibfield  {author} {\bibinfo {author} {\bibfnamefont {Q.-Y.}\ \bibnamefont
  {Lin}}, \bibinfo {author} {\bibfnamefont {X.}~\bibnamefont {Liu}}, \ and\
  \bibinfo {author} {\bibfnamefont {H.-S.}\ \bibnamefont {Xu}},\ }\href
  {\doibase 10.1103/PhysRevD.89.034016} {\bibfield  {journal} {\bibinfo
  {journal} {Phys. Rev. D}\ }\textbf {\bibinfo {volume} {89}},\ \bibinfo
  {pages} {034016} (\bibinfo {year} {2014})},\ \Eprint
  {http://arxiv.org/abs/1312.7073} {arXiv:1312.7073 [hep-ph]} \BibitemShut
  {NoStop}%
\bibitem [{\citenamefont {Wang}\ \emph
  {et~al.}(2015{\natexlab{b}})\citenamefont {Wang}, \citenamefont {Chen},\ and\
  \citenamefont {Guskov}}]{Wang:2015lwa}%
  \BibitemOpen
  \bibfield  {author} {\bibinfo {author} {\bibfnamefont {X.-Y.}\ \bibnamefont
  {Wang}}, \bibinfo {author} {\bibfnamefont {X.-R.}\ \bibnamefont {Chen}}, \
  and\ \bibinfo {author} {\bibfnamefont {A.}~\bibnamefont {Guskov}},\ }\href
  {\doibase 10.1103/PhysRevD.92.094017} {\bibfield  {journal} {\bibinfo
  {journal} {Phys. Rev. D}\ }\textbf {\bibinfo {volume} {92}},\ \bibinfo
  {pages} {094017} (\bibinfo {year} {2015}{\natexlab{b}})},\ \Eprint
  {http://arxiv.org/abs/1503.02125} {arXiv:1503.02125 [hep-ph]} \BibitemShut
  {NoStop}%
\bibitem [{\citenamefont {Albaladejo}\ \emph {et~al.}(2020)\citenamefont
  {Albaladejo}, \citenamefont {Blin}, \citenamefont {Pilloni}, \citenamefont
  {Winney}, \citenamefont {Fern\'andez-Ram\'\i{}rez}, \citenamefont {Mathieu},\
  and\ \citenamefont {Szczepaniak}}]{Albaladejo:2020tzt}%
  \BibitemOpen
  \bibfield  {author} {\bibinfo {author} {\bibfnamefont {M.}~\bibnamefont
  {Albaladejo}}, \bibinfo {author} {\bibfnamefont {A.~N.~H.}\ \bibnamefont
  {Blin}}, \bibinfo {author} {\bibfnamefont {A.}~\bibnamefont {Pilloni}},
  \bibinfo {author} {\bibfnamefont {D.}~\bibnamefont {Winney}}, \bibinfo
  {author} {\bibfnamefont {C.}~\bibnamefont {Fern\'andez-Ram\'\i{}rez}},
  \bibinfo {author} {\bibfnamefont {V.}~\bibnamefont {Mathieu}}, \ and\
  \bibinfo {author} {\bibfnamefont {A.}~\bibnamefont {Szczepaniak}} (\bibinfo
  {collaboration} {JPAC}),\ }\href {\doibase 10.1103/PhysRevD.102.114010}
  {\bibfield  {journal} {\bibinfo  {journal} {Phys. Rev. D}\ }\textbf {\bibinfo
  {volume} {102}},\ \bibinfo {pages} {114010} (\bibinfo {year} {2020})},\
  \Eprint {http://arxiv.org/abs/2008.01001} {arXiv:2008.01001 [hep-ph]}
  \BibitemShut {NoStop}%
\bibitem [{\citenamefont {Du}\ \emph {et~al.}(2020{\natexlab{a}})\citenamefont
  {Du}, \citenamefont {Baru}, \citenamefont {Guo}, \citenamefont {Hanhart},
  \citenamefont {Mei\ss{}ner}, \citenamefont {Nefediev},\ and\ \citenamefont
  {Strakovsky}}]{Du:2020bqj}%
  \BibitemOpen
  \bibfield  {author} {\bibinfo {author} {\bibfnamefont {M.-L.}\ \bibnamefont
  {Du}}, \bibinfo {author} {\bibfnamefont {V.}~\bibnamefont {Baru}}, \bibinfo
  {author} {\bibfnamefont {F.-K.}\ \bibnamefont {Guo}}, \bibinfo {author}
  {\bibfnamefont {C.}~\bibnamefont {Hanhart}}, \bibinfo {author} {\bibfnamefont
  {U.-G.}\ \bibnamefont {Mei\ss{}ner}}, \bibinfo {author} {\bibfnamefont
  {A.}~\bibnamefont {Nefediev}}, \ and\ \bibinfo {author} {\bibfnamefont
  {I.}~\bibnamefont {Strakovsky}},\ }\href {\doibase
  10.1140/epjc/s10052-020-08620-5} {\bibfield  {journal} {\bibinfo  {journal}
  {Eur. Phys. J. C}\ }\textbf {\bibinfo {volume} {80}},\ \bibinfo {pages}
  {1053} (\bibinfo {year} {2020}{\natexlab{a}})},\ \Eprint
  {http://arxiv.org/abs/2009.08345} {arXiv:2009.08345 [hep-ph]} \BibitemShut
  {NoStop}%
\bibitem [{\citenamefont {Xu}\ \emph {et~al.}(2021)\citenamefont {Xu},
  \citenamefont {Chen}, \citenamefont {Yao}, \citenamefont {Binosi},
  \citenamefont {Cui},\ and\ \citenamefont {Roberts}}]{Xu:2021mju}%
  \BibitemOpen
  \bibfield  {author} {\bibinfo {author} {\bibfnamefont {Y.-Z.}\ \bibnamefont
  {Xu}}, \bibinfo {author} {\bibfnamefont {S.}~\bibnamefont {Chen}}, \bibinfo
  {author} {\bibfnamefont {Z.-Q.}\ \bibnamefont {Yao}}, \bibinfo {author}
  {\bibfnamefont {D.}~\bibnamefont {Binosi}}, \bibinfo {author} {\bibfnamefont
  {Z.-F.}\ \bibnamefont {Cui}}, \ and\ \bibinfo {author} {\bibfnamefont
  {C.~D.}\ \bibnamefont {Roberts}},\ }\href {\doibase
  10.1140/epjc/s10052-021-09673-w} {\bibfield  {journal} {\bibinfo  {journal}
  {Eur. Phys. J. C}\ }\textbf {\bibinfo {volume} {81}},\ \bibinfo {pages} {895}
  (\bibinfo {year} {2021})},\ \Eprint {http://arxiv.org/abs/2107.03488}
  {arXiv:2107.03488 [hep-ph]} \BibitemShut {NoStop}%
\bibitem [{\citenamefont {Yang}\ and\ \citenamefont
  {Guo}(2021)}]{Yang:2021jof}%
  \BibitemOpen
  \bibfield  {author} {\bibinfo {author} {\bibfnamefont {Z.}~\bibnamefont
  {Yang}}\ and\ \bibinfo {author} {\bibfnamefont {F.-K.}\ \bibnamefont {Guo}},\
  }\href {\doibase 10.1088/1674-1137/ac2359} {\bibfield  {journal} {\bibinfo
  {journal} {Chin. Phys. C}\ }\textbf {\bibinfo {volume} {45}},\ \bibinfo
  {pages} {123101} (\bibinfo {year} {2021})},\ \Eprint
  {http://arxiv.org/abs/2107.12247} {arXiv:2107.12247 [hep-ph]} \BibitemShut
  {NoStop}%
\bibitem [{\citenamefont {Artoisenet}\ and\ \citenamefont
  {Braaten}(2011)}]{Artoisenet:2010uu}%
  \BibitemOpen
  \bibfield  {author} {\bibinfo {author} {\bibfnamefont {P.}~\bibnamefont
  {Artoisenet}}\ and\ \bibinfo {author} {\bibfnamefont {E.}~\bibnamefont
  {Braaten}},\ }\href {\doibase 10.1103/PhysRevD.83.014019} {\bibfield
  {journal} {\bibinfo  {journal} {Phys. Rev. D}\ }\textbf {\bibinfo {volume}
  {83}},\ \bibinfo {pages} {014019} (\bibinfo {year} {2011})},\ \Eprint
  {http://arxiv.org/abs/1007.2868} {arXiv:1007.2868 [hep-ph]} \BibitemShut
  {NoStop}%
\bibitem [{\citenamefont {Guo}\ \emph {et~al.}(2014{\natexlab{a}})\citenamefont
  {Guo}, \citenamefont {Mei\ss{}ner}, \citenamefont {Wang},\ and\ \citenamefont
  {Yang}}]{Guo:2014ppa}%
  \BibitemOpen
  \bibfield  {author} {\bibinfo {author} {\bibfnamefont {F.-K.}\ \bibnamefont
  {Guo}}, \bibinfo {author} {\bibfnamefont {U.-G.}\ \bibnamefont
  {Mei\ss{}ner}}, \bibinfo {author} {\bibfnamefont {W.}~\bibnamefont {Wang}}, \
  and\ \bibinfo {author} {\bibfnamefont {Z.}~\bibnamefont {Yang}},\ }\href
  {\doibase 10.1007/JHEP05(2014)138} {\bibfield  {journal} {\bibinfo  {journal}
  {JHEP}\ }\textbf {\bibinfo {volume} {05}},\ \bibinfo {pages} {138} (\bibinfo
  {year} {2014}{\natexlab{a}})},\ \Eprint {http://arxiv.org/abs/1403.4032}
  {arXiv:1403.4032 [hep-ph]} \BibitemShut {NoStop}%
\bibitem [{\citenamefont {Guo}\ \emph {et~al.}(2014{\natexlab{b}})\citenamefont
  {Guo}, \citenamefont {Mei\ss{}ner}, \citenamefont {Wang},\ and\ \citenamefont
  {Yang}}]{Guo:2014sca}%
  \BibitemOpen
  \bibfield  {author} {\bibinfo {author} {\bibfnamefont {F.-K.}\ \bibnamefont
  {Guo}}, \bibinfo {author} {\bibfnamefont {U.-G.}\ \bibnamefont
  {Mei\ss{}ner}}, \bibinfo {author} {\bibfnamefont {W.}~\bibnamefont {Wang}}, \
  and\ \bibinfo {author} {\bibfnamefont {Z.}~\bibnamefont {Yang}},\ }\href
  {\doibase 10.1140/epjc/s10052-014-3063-4} {\bibfield  {journal} {\bibinfo
  {journal} {Eur. Phys. J. C}\ }\textbf {\bibinfo {volume} {74}},\ \bibinfo
  {pages} {3063} (\bibinfo {year} {2014}{\natexlab{b}})},\ \Eprint
  {http://arxiv.org/abs/1402.6236} {arXiv:1402.6236 [hep-ph]} \BibitemShut
  {NoStop}%
\bibitem [{\citenamefont {Bignamini}\ \emph {et~al.}(2009)\citenamefont
  {Bignamini}, \citenamefont {Grinstein}, \citenamefont {Piccinini},
  \citenamefont {Polosa},\ and\ \citenamefont {Sabelli}}]{Bignamini:2009sk}%
  \BibitemOpen
  \bibfield  {author} {\bibinfo {author} {\bibfnamefont {C.}~\bibnamefont
  {Bignamini}}, \bibinfo {author} {\bibfnamefont {B.}~\bibnamefont
  {Grinstein}}, \bibinfo {author} {\bibfnamefont {F.}~\bibnamefont
  {Piccinini}}, \bibinfo {author} {\bibfnamefont {A.~D.}\ \bibnamefont
  {Polosa}}, \ and\ \bibinfo {author} {\bibfnamefont {C.}~\bibnamefont
  {Sabelli}},\ }\href {\doibase 10.1103/PhysRevLett.103.162001} {\bibfield
  {journal} {\bibinfo  {journal} {Phys. Rev. Lett.}\ }\textbf {\bibinfo
  {volume} {103}},\ \bibinfo {pages} {162001} (\bibinfo {year} {2009})},\
  \Eprint {http://arxiv.org/abs/0906.0882} {arXiv:0906.0882 [hep-ph]}
  \BibitemShut {NoStop}%
\bibitem [{\citenamefont {Artoisenet}\ and\ \citenamefont
  {Braaten}(2010)}]{Artoisenet:2009wk}%
  \BibitemOpen
  \bibfield  {author} {\bibinfo {author} {\bibfnamefont {P.}~\bibnamefont
  {Artoisenet}}\ and\ \bibinfo {author} {\bibfnamefont {E.}~\bibnamefont
  {Braaten}},\ }\href {\doibase 10.1103/PhysRevD.81.114018} {\bibfield
  {journal} {\bibinfo  {journal} {Phys. Rev. D}\ }\textbf {\bibinfo {volume}
  {81}},\ \bibinfo {pages} {114018} (\bibinfo {year} {2010})},\ \Eprint
  {http://arxiv.org/abs/0911.2016} {arXiv:0911.2016 [hep-ph]} \BibitemShut
  {NoStop}%
\bibitem [{\citenamefont {Albaladejo}\ \emph {et~al.}(2017)\citenamefont
  {Albaladejo}, \citenamefont {Guo}, \citenamefont {Hanhart}, \citenamefont
  {Mei\ss{}ner}, \citenamefont {Nieves}, \citenamefont {Nogga},\ and\
  \citenamefont {Yang}}]{Albaladejo:2017blx}%
  \BibitemOpen
  \bibfield  {author} {\bibinfo {author} {\bibfnamefont {M.}~\bibnamefont
  {Albaladejo}}, \bibinfo {author} {\bibfnamefont {F.-K.}\ \bibnamefont {Guo}},
  \bibinfo {author} {\bibfnamefont {C.}~\bibnamefont {Hanhart}}, \bibinfo
  {author} {\bibfnamefont {U.-G.}\ \bibnamefont {Mei\ss{}ner}}, \bibinfo
  {author} {\bibfnamefont {J.}~\bibnamefont {Nieves}}, \bibinfo {author}
  {\bibfnamefont {A.}~\bibnamefont {Nogga}}, \ and\ \bibinfo {author}
  {\bibfnamefont {Z.}~\bibnamefont {Yang}},\ }\href {\doibase
  10.1088/1674-1137/41/12/121001} {\bibfield  {journal} {\bibinfo  {journal}
  {Chin. Phys. C}\ }\textbf {\bibinfo {volume} {41}},\ \bibinfo {pages}
  {121001} (\bibinfo {year} {2017})},\ \Eprint
  {http://arxiv.org/abs/1709.09101} {arXiv:1709.09101 [hep-ph]} \BibitemShut
  {NoStop}%
\bibitem [{\citenamefont {Guo}\ \emph {et~al.}(2014{\natexlab{c}})\citenamefont
  {Guo}, \citenamefont {Mei\ss{}ner},\ and\ \citenamefont
  {Wang}}]{Guo:2013ufa}%
  \BibitemOpen
  \bibfield  {author} {\bibinfo {author} {\bibfnamefont {F.-K.}\ \bibnamefont
  {Guo}}, \bibinfo {author} {\bibfnamefont {U.-G.}\ \bibnamefont
  {Mei\ss{}ner}}, \ and\ \bibinfo {author} {\bibfnamefont {W.}~\bibnamefont
  {Wang}},\ }\href {\doibase 10.1088/0253-6102/61/3/14} {\bibfield  {journal}
  {\bibinfo  {journal} {Commun. Theor. Phys.}\ }\textbf {\bibinfo {volume}
  {61}},\ \bibinfo {pages} {354} (\bibinfo {year} {2014}{\natexlab{c}})},\
  \Eprint {http://arxiv.org/abs/1308.0193} {arXiv:1308.0193 [hep-ph]}
  \BibitemShut {NoStop}%
\bibitem [{\citenamefont {Ling}\ \emph {et~al.}(2021)\citenamefont {Ling},
  \citenamefont {Dai}, \citenamefont {Du},\ and\ \citenamefont
  {Wang}}]{Ling:2021sld}%
  \BibitemOpen
  \bibfield  {author} {\bibinfo {author} {\bibfnamefont {P.}~\bibnamefont
  {Ling}}, \bibinfo {author} {\bibfnamefont {X.-H.}\ \bibnamefont {Dai}},
  \bibinfo {author} {\bibfnamefont {M.-L.}\ \bibnamefont {Du}}, \ and\ \bibinfo
  {author} {\bibfnamefont {Q.}~\bibnamefont {Wang}},\ }\href {\doibase
  10.1140/epjc/s10052-021-09613-8} {\bibfield  {journal} {\bibinfo  {journal}
  {Eur. Phys. J. C}\ }\textbf {\bibinfo {volume} {81}},\ \bibinfo {pages} {819}
  (\bibinfo {year} {2021})},\ \Eprint {http://arxiv.org/abs/2104.11133}
  {arXiv:2104.11133 [hep-ph]} \BibitemShut {NoStop}%
\bibitem [{\citenamefont {Shi}\ \emph {et~al.}(2022)\citenamefont {Shi},
  \citenamefont {Zhang}, \citenamefont {Guo},\ and\ \citenamefont
  {Yang}}]{Shi:2021hzm}%
  \BibitemOpen
  \bibfield  {author} {\bibinfo {author} {\bibfnamefont {P.-P.}\ \bibnamefont
  {Shi}}, \bibinfo {author} {\bibfnamefont {Z.-H.}\ \bibnamefont {Zhang}},
  \bibinfo {author} {\bibfnamefont {F.-K.}\ \bibnamefont {Guo}}, \ and\
  \bibinfo {author} {\bibfnamefont {Z.}~\bibnamefont {Yang}},\ }\href {\doibase
  10.1103/PhysRevD.105.034024} {\bibfield  {journal} {\bibinfo  {journal}
  {Phys. Rev. D}\ }\textbf {\bibinfo {volume} {105}},\ \bibinfo {pages}
  {034024} (\bibinfo {year} {2022})},\ \Eprint
  {http://arxiv.org/abs/2111.13496} {arXiv:2111.13496 [hep-ph]} \BibitemShut
  {NoStop}%
\bibitem [{\citenamefont {Bauer}(2005)}]{Bauer:2004bc}%
  \BibitemOpen
  \bibfield  {author} {\bibinfo {author} {\bibfnamefont {G.}~\bibnamefont
  {Bauer}} (\bibinfo {collaboration} {CDF}),\ }\href {\doibase
  10.1142/S0217751X05027552} {\bibfield  {journal} {\bibinfo  {journal} {Int.
  J. Mod. Phys. A}\ }\textbf {\bibinfo {volume} {20}},\ \bibinfo {pages} {3765}
  (\bibinfo {year} {2005})},\ \Eprint {http://arxiv.org/abs/hep-ex/0409052}
  {arXiv:hep-ex/0409052} \BibitemShut {NoStop}%
\bibitem [{\citenamefont {Chatrchyan}\ \emph {et~al.}(2013)\citenamefont
  {Chatrchyan} \emph {et~al.}}]{CMS:2013fpt}%
  \BibitemOpen
  \bibfield  {author} {\bibinfo {author} {\bibfnamefont {S.}~\bibnamefont
  {Chatrchyan}} \emph {et~al.} (\bibinfo {collaboration} {CMS}),\ }\href
  {\doibase 10.1007/JHEP04(2013)154} {\bibfield  {journal} {\bibinfo  {journal}
  {JHEP}\ }\textbf {\bibinfo {volume} {04}},\ \bibinfo {pages} {154} (\bibinfo
  {year} {2013})},\ \Eprint {http://arxiv.org/abs/1302.3968} {arXiv:1302.3968
  [hep-ex]} \BibitemShut {NoStop}%
\bibitem [{\citenamefont {Aaij}\ \emph {et~al.}(2019)\citenamefont {Aaij} \emph
  {et~al.}}]{LHCb:2019kea}%
  \BibitemOpen
  \bibfield  {author} {\bibinfo {author} {\bibfnamefont {R.}~\bibnamefont
  {Aaij}} \emph {et~al.} (\bibinfo {collaboration} {LHCb}),\ }\href {\doibase
  10.1103/PhysRevLett.122.222001} {\bibfield  {journal} {\bibinfo  {journal}
  {Phys. Rev. Lett.}\ }\textbf {\bibinfo {volume} {122}},\ \bibinfo {pages}
  {222001} (\bibinfo {year} {2019})},\ \Eprint
  {http://arxiv.org/abs/1904.03947} {arXiv:1904.03947 [hep-ex]} \BibitemShut
  {NoStop}%
\bibitem [{\citenamefont {Aaij}\ \emph {et~al.}(2021)\citenamefont {Aaij} \emph
  {et~al.}}]{LHCb:2020jpq}%
  \BibitemOpen
  \bibfield  {author} {\bibinfo {author} {\bibfnamefont {R.}~\bibnamefont
  {Aaij}} \emph {et~al.} (\bibinfo {collaboration} {LHCb}),\ }\href {\doibase
  10.1016/j.scib.2021.02.030} {\bibfield  {journal} {\bibinfo  {journal} {Sci.
  Bull.}\ }\textbf {\bibinfo {volume} {66}},\ \bibinfo {pages} {1278} (\bibinfo
  {year} {2021})},\ \Eprint {http://arxiv.org/abs/2012.10380} {arXiv:2012.10380
  [hep-ex]} \BibitemShut {NoStop}%
\bibitem [{\citenamefont {Aaij}\ \emph
  {et~al.}(2022{\natexlab{a}})\citenamefont {Aaij} \emph
  {et~al.}}]{LHCb:2021auc}%
  \BibitemOpen
  \bibfield  {author} {\bibinfo {author} {\bibfnamefont {R.}~\bibnamefont
  {Aaij}} \emph {et~al.} (\bibinfo {collaboration} {LHCb}),\ }\href {\doibase
  10.1038/s41467-022-30206-w} {\bibfield  {journal} {\bibinfo  {journal}
  {Nature Commun.}\ }\textbf {\bibinfo {volume} {13}},\ \bibinfo {pages} {3351}
  (\bibinfo {year} {2022}{\natexlab{a}})},\ \Eprint
  {http://arxiv.org/abs/2109.01056} {arXiv:2109.01056 [hep-ex]} \BibitemShut
  {NoStop}%
\bibitem [{\citenamefont {Aaij}\ \emph
  {et~al.}(2022{\natexlab{b}})\citenamefont {Aaij} \emph
  {et~al.}}]{LHCb:2021vvq}%
  \BibitemOpen
  \bibfield  {author} {\bibinfo {author} {\bibfnamefont {R.}~\bibnamefont
  {Aaij}} \emph {et~al.} (\bibinfo {collaboration} {LHCb}),\ }\href {\doibase
  10.1038/s41567-022-01614-y} {\bibfield  {journal} {\bibinfo  {journal}
  {Nature Phys.}\ }\textbf {\bibinfo {volume} {18}},\ \bibinfo {pages} {751}
  (\bibinfo {year} {2022}{\natexlab{b}})},\ \Eprint
  {http://arxiv.org/abs/2109.01038} {arXiv:2109.01038 [hep-ex]} \BibitemShut
  {NoStop}%
\bibitem [{\citenamefont {Choi}\ \emph {et~al.}(2003)\citenamefont {Choi} \emph
  {et~al.}}]{Belle:2003nnu}%
  \BibitemOpen
  \bibfield  {author} {\bibinfo {author} {\bibfnamefont {S.~K.}\ \bibnamefont
  {Choi}} \emph {et~al.} (\bibinfo {collaboration} {Belle}),\ }\href {\doibase
  10.1103/PhysRevLett.91.262001} {\bibfield  {journal} {\bibinfo  {journal}
  {Phys. Rev. Lett.}\ }\textbf {\bibinfo {volume} {91}},\ \bibinfo {pages}
  {262001} (\bibinfo {year} {2003})},\ \Eprint
  {http://arxiv.org/abs/hep-ex/0309032} {arXiv:hep-ex/0309032} \BibitemShut
  {NoStop}%
\bibitem [{\citenamefont {Ablikim}\ \emph {et~al.}(2013)\citenamefont {Ablikim}
  \emph {et~al.}}]{BESIII:2013ris}%
  \BibitemOpen
  \bibfield  {author} {\bibinfo {author} {\bibfnamefont {M.}~\bibnamefont
  {Ablikim}} \emph {et~al.} (\bibinfo {collaboration} {BESIII}),\ }\href
  {\doibase 10.1103/PhysRevLett.110.252001} {\bibfield  {journal} {\bibinfo
  {journal} {Phys. Rev. Lett.}\ }\textbf {\bibinfo {volume} {110}},\ \bibinfo
  {pages} {252001} (\bibinfo {year} {2013})},\ \Eprint
  {http://arxiv.org/abs/1303.5949} {arXiv:1303.5949 [hep-ex]} \BibitemShut
  {NoStop}%
\bibitem [{\citenamefont {Liu}\ \emph {et~al.}(2013)\citenamefont {Liu} \emph
  {et~al.}}]{Belle:2013yex}%
  \BibitemOpen
  \bibfield  {author} {\bibinfo {author} {\bibfnamefont {Z.~Q.}\ \bibnamefont
  {Liu}} \emph {et~al.} (\bibinfo {collaboration} {Belle}),\ }\href {\doibase
  10.1103/PhysRevLett.110.252002} {\bibfield  {journal} {\bibinfo  {journal}
  {Phys. Rev. Lett.}\ }\textbf {\bibinfo {volume} {110}},\ \bibinfo {pages}
  {252002} (\bibinfo {year} {2013})},\ \bibinfo {note} {[Erratum:
  Phys.Rev.Lett. 111, 019901 (2013)]},\ \Eprint
  {http://arxiv.org/abs/1304.0121} {arXiv:1304.0121 [hep-ex]} \BibitemShut
  {NoStop}%
\bibitem [{\citenamefont {Ablikim}\ \emph {et~al.}(2015)\citenamefont {Ablikim}
  \emph {et~al.}}]{BESIII:2015cld}%
  \BibitemOpen
  \bibfield  {author} {\bibinfo {author} {\bibfnamefont {M.}~\bibnamefont
  {Ablikim}} \emph {et~al.} (\bibinfo {collaboration} {BESIII}),\ }\href
  {\doibase 10.1103/PhysRevLett.115.112003} {\bibfield  {journal} {\bibinfo
  {journal} {Phys. Rev. Lett.}\ }\textbf {\bibinfo {volume} {115}},\ \bibinfo
  {pages} {112003} (\bibinfo {year} {2015})},\ \Eprint
  {http://arxiv.org/abs/1506.06018} {arXiv:1506.06018 [hep-ex]} \BibitemShut
  {NoStop}%
\bibitem [{\citenamefont {Dong}\ \emph
  {et~al.}(2021{\natexlab{a}})\citenamefont {Dong}, \citenamefont {Guo},\ and\
  \citenamefont {Zou}}]{Dong:2021juy}%
  \BibitemOpen
  \bibfield  {author} {\bibinfo {author} {\bibfnamefont {X.-K.}\ \bibnamefont
  {Dong}}, \bibinfo {author} {\bibfnamefont {F.-K.}\ \bibnamefont {Guo}}, \
  and\ \bibinfo {author} {\bibfnamefont {B.-S.}\ \bibnamefont {Zou}},\ }\href
  {\doibase 10.13725/j.cnki.pip.2021.02.001} {\bibfield  {journal} {\bibinfo
  {journal} {Progr. Phys.}\ }\textbf {\bibinfo {volume} {41}},\ \bibinfo
  {pages} {65} (\bibinfo {year} {2021}{\natexlab{a}})},\ \Eprint
  {http://arxiv.org/abs/2101.01021} {arXiv:2101.01021 [hep-ph]} \BibitemShut
  {NoStop}%
\bibitem [{\citenamefont {Dong}\ \emph
  {et~al.}(2021{\natexlab{b}})\citenamefont {Dong}, \citenamefont {Guo},\ and\
  \citenamefont {Zou}}]{Dong:2021bvy}%
  \BibitemOpen
  \bibfield  {author} {\bibinfo {author} {\bibfnamefont {X.-K.}\ \bibnamefont
  {Dong}}, \bibinfo {author} {\bibfnamefont {F.-K.}\ \bibnamefont {Guo}}, \
  and\ \bibinfo {author} {\bibfnamefont {B.-S.}\ \bibnamefont {Zou}},\ }\href
  {\doibase 10.1088/1572-9494/ac27a2} {\bibfield  {journal} {\bibinfo
  {journal} {Commun. Theor. Phys.}\ }\textbf {\bibinfo {volume} {73}},\
  \bibinfo {pages} {125201} (\bibinfo {year} {2021}{\natexlab{b}})},\ \Eprint
  {http://arxiv.org/abs/2108.02673} {arXiv:2108.02673 [hep-ph]} \BibitemShut
  {NoStop}%
\bibitem [{\citenamefont {Braaten}\ and\ \citenamefont
  {Kusunoki}(2005)}]{Braaten:2005jj}%
  \BibitemOpen
  \bibfield  {author} {\bibinfo {author} {\bibfnamefont {E.}~\bibnamefont
  {Braaten}}\ and\ \bibinfo {author} {\bibfnamefont {M.}~\bibnamefont
  {Kusunoki}},\ }\href {\doibase 10.1103/PhysRevD.72.014012} {\bibfield
  {journal} {\bibinfo  {journal} {Phys. Rev. D}\ }\textbf {\bibinfo {volume}
  {72}},\ \bibinfo {pages} {014012} (\bibinfo {year} {2005})},\ \Eprint
  {http://arxiv.org/abs/hep-ph/0506087} {arXiv:hep-ph/0506087} \BibitemShut
  {NoStop}%
\bibitem [{\citenamefont {Sjostrand}\ \emph {et~al.}(2006)\citenamefont
  {Sjostrand}, \citenamefont {Mrenna},\ and\ \citenamefont
  {Skands}}]{Sjostrand:2006za}%
  \BibitemOpen
  \bibfield  {author} {\bibinfo {author} {\bibfnamefont {T.}~\bibnamefont
  {Sjostrand}}, \bibinfo {author} {\bibfnamefont {S.}~\bibnamefont {Mrenna}}, \
  and\ \bibinfo {author} {\bibfnamefont {P.~Z.}\ \bibnamefont {Skands}},\
  }\href {\doibase 10.1088/1126-6708/2006/05/026} {\bibfield  {journal}
  {\bibinfo  {journal} {JHEP}\ }\textbf {\bibinfo {volume} {05}},\ \bibinfo
  {pages} {026} (\bibinfo {year} {2006})},\ \Eprint
  {http://arxiv.org/abs/hep-ph/0603175} {arXiv:hep-ph/0603175} \BibitemShut
  {NoStop}%
\bibitem [{\citenamefont {Nieves}\ and\ \citenamefont
  {Valderrama}(2012)}]{Nieves:2012tt}%
  \BibitemOpen
  \bibfield  {author} {\bibinfo {author} {\bibfnamefont {J.}~\bibnamefont
  {Nieves}}\ and\ \bibinfo {author} {\bibfnamefont {M.~P.}\ \bibnamefont
  {Valderrama}},\ }\href {\doibase 10.1103/PhysRevD.86.056004} {\bibfield
  {journal} {\bibinfo  {journal} {Phys. Rev. D}\ }\textbf {\bibinfo {volume}
  {86}},\ \bibinfo {pages} {056004} (\bibinfo {year} {2012})},\ \Eprint
  {http://arxiv.org/abs/1204.2790} {arXiv:1204.2790 [hep-ph]} \BibitemShut
  {NoStop}%
\bibitem [{\citenamefont {Guo}\ \emph {et~al.}(2013)\citenamefont {Guo},
  \citenamefont {Hidalgo-Duque}, \citenamefont {Nieves},\ and\ \citenamefont
  {Valderrama}}]{Guo:2013sya}%
  \BibitemOpen
  \bibfield  {author} {\bibinfo {author} {\bibfnamefont {F.-K.}\ \bibnamefont
  {Guo}}, \bibinfo {author} {\bibfnamefont {C.}~\bibnamefont {Hidalgo-Duque}},
  \bibinfo {author} {\bibfnamefont {J.}~\bibnamefont {Nieves}}, \ and\ \bibinfo
  {author} {\bibfnamefont {M.~P.}\ \bibnamefont {Valderrama}},\ }\href
  {\doibase 10.1103/PhysRevD.88.054007} {\bibfield  {journal} {\bibinfo
  {journal} {Phys. Rev. D}\ }\textbf {\bibinfo {volume} {88}},\ \bibinfo
  {pages} {054007} (\bibinfo {year} {2013})},\ \Eprint
  {http://arxiv.org/abs/1303.6608} {arXiv:1303.6608 [hep-ph]} \BibitemShut
  {NoStop}%
\bibitem [{\citenamefont {Liu}\ \emph {et~al.}(2019{\natexlab{b}})\citenamefont
  {Liu}, \citenamefont {Pan}, \citenamefont {Peng}, \citenamefont
  {S\'anchez~S\'anchez}, \citenamefont {Geng}, \citenamefont {Hosaka},\ and\
  \citenamefont {Pavon~Valderrama}}]{Liu:2019tjn}%
  \BibitemOpen
  \bibfield  {author} {\bibinfo {author} {\bibfnamefont {M.-Z.}\ \bibnamefont
  {Liu}}, \bibinfo {author} {\bibfnamefont {Y.-W.}\ \bibnamefont {Pan}},
  \bibinfo {author} {\bibfnamefont {F.-Z.}\ \bibnamefont {Peng}}, \bibinfo
  {author} {\bibfnamefont {M.}~\bibnamefont {S\'anchez~S\'anchez}}, \bibinfo
  {author} {\bibfnamefont {L.-S.}\ \bibnamefont {Geng}}, \bibinfo {author}
  {\bibfnamefont {A.}~\bibnamefont {Hosaka}}, \ and\ \bibinfo {author}
  {\bibfnamefont {M.}~\bibnamefont {Pavon~Valderrama}},\ }\href {\doibase
  10.1103/PhysRevLett.122.242001} {\bibfield  {journal} {\bibinfo  {journal}
  {Phys. Rev. Lett.}\ }\textbf {\bibinfo {volume} {122}},\ \bibinfo {pages}
  {242001} (\bibinfo {year} {2019}{\natexlab{b}})},\ \Eprint
  {http://arxiv.org/abs/1903.11560} {arXiv:1903.11560 [hep-ph]} \BibitemShut
  {NoStop}%
\bibitem [{\citenamefont {Yang}\ \emph {et~al.}(2021)\citenamefont {Yang},
  \citenamefont {Cao}, \citenamefont {Guo}, \citenamefont {Nieves},\ and\
  \citenamefont {Valderrama}}]{Yang:2020nrt}%
  \BibitemOpen
  \bibfield  {author} {\bibinfo {author} {\bibfnamefont {Z.}~\bibnamefont
  {Yang}}, \bibinfo {author} {\bibfnamefont {X.}~\bibnamefont {Cao}}, \bibinfo
  {author} {\bibfnamefont {F.-K.}\ \bibnamefont {Guo}}, \bibinfo {author}
  {\bibfnamefont {J.}~\bibnamefont {Nieves}}, \ and\ \bibinfo {author}
  {\bibfnamefont {M.~P.}\ \bibnamefont {Valderrama}},\ }\href {\doibase
  10.1103/PhysRevD.103.074029} {\bibfield  {journal} {\bibinfo  {journal}
  {Phys. Rev. D}\ }\textbf {\bibinfo {volume} {103}},\ \bibinfo {pages}
  {074029} (\bibinfo {year} {2021})},\ \Eprint
  {http://arxiv.org/abs/2011.08725} {arXiv:2011.08725 [hep-ph]} \BibitemShut
  {NoStop}%
\bibitem [{\citenamefont {Aaij}\ \emph {et~al.}(2020)\citenamefont {Aaij} \emph
  {et~al.}}]{LHCb:2020xds}%
  \BibitemOpen
  \bibfield  {author} {\bibinfo {author} {\bibfnamefont {R.}~\bibnamefont
  {Aaij}} \emph {et~al.} (\bibinfo {collaboration} {LHCb}),\ }\href {\doibase
  10.1103/PhysRevD.102.092005} {\bibfield  {journal} {\bibinfo  {journal}
  {Phys. Rev. D}\ }\textbf {\bibinfo {volume} {102}},\ \bibinfo {pages}
  {092005} (\bibinfo {year} {2020})},\ \Eprint
  {http://arxiv.org/abs/2005.13419} {arXiv:2005.13419 [hep-ex]} \BibitemShut
  {NoStop}%
\bibitem [{\citenamefont {Hanhart}\ \emph {et~al.}(2012)\citenamefont
  {Hanhart}, \citenamefont {Kalashnikova}, \citenamefont {Kudryavtsev},\ and\
  \citenamefont {Nefediev}}]{Hanhart:2011tn}%
  \BibitemOpen
  \bibfield  {author} {\bibinfo {author} {\bibfnamefont {C.}~\bibnamefont
  {Hanhart}}, \bibinfo {author} {\bibfnamefont {Y.~S.}\ \bibnamefont
  {Kalashnikova}}, \bibinfo {author} {\bibfnamefont {A.~E.}\ \bibnamefont
  {Kudryavtsev}}, \ and\ \bibinfo {author} {\bibfnamefont {A.~V.}\ \bibnamefont
  {Nefediev}},\ }\href {\doibase 10.1103/PhysRevD.85.011501} {\bibfield
  {journal} {\bibinfo  {journal} {Phys. Rev. D}\ }\textbf {\bibinfo {volume}
  {85}},\ \bibinfo {pages} {011501} (\bibinfo {year} {2012})},\ \Eprint
  {http://arxiv.org/abs/1111.6241} {arXiv:1111.6241 [hep-ph]} \BibitemShut
  {NoStop}%
\bibitem [{\citenamefont {Hidalgo-Duque}\ \emph {et~al.}(2013)\citenamefont
  {Hidalgo-Duque}, \citenamefont {Nieves},\ and\ \citenamefont
  {Valderrama}}]{Hidalgo-Duque:2012rqv}%
  \BibitemOpen
  \bibfield  {author} {\bibinfo {author} {\bibfnamefont {C.}~\bibnamefont
  {Hidalgo-Duque}}, \bibinfo {author} {\bibfnamefont {J.}~\bibnamefont
  {Nieves}}, \ and\ \bibinfo {author} {\bibfnamefont {M.~P.}\ \bibnamefont
  {Valderrama}},\ }\href {\doibase 10.1103/PhysRevD.87.076006} {\bibfield
  {journal} {\bibinfo  {journal} {Phys. Rev. D}\ }\textbf {\bibinfo {volume}
  {87}},\ \bibinfo {pages} {076006} (\bibinfo {year} {2013})},\ \Eprint
  {http://arxiv.org/abs/1210.5431} {arXiv:1210.5431 [hep-ph]} \BibitemShut
  {NoStop}%
\bibitem [{\citenamefont {Ablikim}\ \emph {et~al.}(2021)\citenamefont {Ablikim}
  \emph {et~al.}}]{BESIII:2020qkh}%
  \BibitemOpen
  \bibfield  {author} {\bibinfo {author} {\bibfnamefont {M.}~\bibnamefont
  {Ablikim}} \emph {et~al.} (\bibinfo {collaboration} {BESIII}),\ }\href
  {\doibase 10.1103/PhysRevLett.126.102001} {\bibfield  {journal} {\bibinfo
  {journal} {Phys. Rev. Lett.}\ }\textbf {\bibinfo {volume} {126}},\ \bibinfo
  {pages} {102001} (\bibinfo {year} {2021})},\ \Eprint
  {http://arxiv.org/abs/2011.07855} {arXiv:2011.07855 [hep-ex]} \BibitemShut
  {NoStop}%
\bibitem [{\citenamefont {Du}\ \emph {et~al.}(2022)\citenamefont {Du},
  \citenamefont {Baru}, \citenamefont {Dong}, \citenamefont {Filin},
  \citenamefont {Guo}, \citenamefont {Hanhart}, \citenamefont {Nefediev},
  \citenamefont {Nieves},\ and\ \citenamefont {Wang}}]{Du:2021zzh}%
  \BibitemOpen
  \bibfield  {author} {\bibinfo {author} {\bibfnamefont {M.-L.}\ \bibnamefont
  {Du}}, \bibinfo {author} {\bibfnamefont {V.}~\bibnamefont {Baru}}, \bibinfo
  {author} {\bibfnamefont {X.-K.}\ \bibnamefont {Dong}}, \bibinfo {author}
  {\bibfnamefont {A.}~\bibnamefont {Filin}}, \bibinfo {author} {\bibfnamefont
  {F.-K.}\ \bibnamefont {Guo}}, \bibinfo {author} {\bibfnamefont
  {C.}~\bibnamefont {Hanhart}}, \bibinfo {author} {\bibfnamefont
  {A.}~\bibnamefont {Nefediev}}, \bibinfo {author} {\bibfnamefont
  {J.}~\bibnamefont {Nieves}}, \ and\ \bibinfo {author} {\bibfnamefont
  {Q.}~\bibnamefont {Wang}},\ }\href {\doibase 10.1103/PhysRevD.105.014024}
  {\bibfield  {journal} {\bibinfo  {journal} {Phys. Rev. D}\ }\textbf {\bibinfo
  {volume} {105}},\ \bibinfo {pages} {014024} (\bibinfo {year} {2022})},\
  \Eprint {http://arxiv.org/abs/2110.13765} {arXiv:2110.13765 [hep-ph]}
  \BibitemShut {NoStop}%
\bibitem [{\citenamefont {Du}\ \emph {et~al.}(2021{\natexlab{a}})\citenamefont
  {Du}, \citenamefont {Guo},\ and\ \citenamefont {Oller}}]{Du:2021bgb}%
  \BibitemOpen
  \bibfield  {author} {\bibinfo {author} {\bibfnamefont {M.-L.}\ \bibnamefont
  {Du}}, \bibinfo {author} {\bibfnamefont {Z.-H.}\ \bibnamefont {Guo}}, \ and\
  \bibinfo {author} {\bibfnamefont {J.~A.}\ \bibnamefont {Oller}},\ }\href
  {\doibase 10.1103/PhysRevD.104.114034} {\bibfield  {journal} {\bibinfo
  {journal} {Phys. Rev. D}\ }\textbf {\bibinfo {volume} {104}},\ \bibinfo
  {pages} {114034} (\bibinfo {year} {2021}{\natexlab{a}})},\ \Eprint
  {http://arxiv.org/abs/2109.14237} {arXiv:2109.14237 [hep-ph]} \BibitemShut
  {NoStop}%
\bibitem [{\citenamefont {Liu}\ \emph {et~al.}(2021)\citenamefont {Liu},
  \citenamefont {Pan},\ and\ \citenamefont {Geng}}]{Liu:2020hcv}%
  \BibitemOpen
  \bibfield  {author} {\bibinfo {author} {\bibfnamefont {M.-Z.}\ \bibnamefont
  {Liu}}, \bibinfo {author} {\bibfnamefont {Y.-W.}\ \bibnamefont {Pan}}, \ and\
  \bibinfo {author} {\bibfnamefont {L.-S.}\ \bibnamefont {Geng}},\ }\href
  {\doibase 10.1103/PhysRevD.103.034003} {\bibfield  {journal} {\bibinfo
  {journal} {Phys. Rev. D}\ }\textbf {\bibinfo {volume} {103}},\ \bibinfo
  {pages} {034003} (\bibinfo {year} {2021})},\ \Eprint
  {http://arxiv.org/abs/2011.07935} {arXiv:2011.07935 [hep-ph]} \BibitemShut
  {NoStop}%
\bibitem [{\citenamefont {Chen}(2021)}]{Chen:2020kco}%
  \BibitemOpen
  \bibfield  {author} {\bibinfo {author} {\bibfnamefont {R.}~\bibnamefont
  {Chen}},\ }\href {\doibase 10.1103/PhysRevD.103.054007} {\bibfield  {journal}
  {\bibinfo  {journal} {Phys. Rev. D}\ }\textbf {\bibinfo {volume} {103}},\
  \bibinfo {pages} {054007} (\bibinfo {year} {2021})},\ \Eprint
  {http://arxiv.org/abs/2011.07214} {arXiv:2011.07214 [hep-ph]} \BibitemShut
  {NoStop}%
\bibitem [{\citenamefont {Peng}\ \emph {et~al.}(2021)\citenamefont {Peng},
  \citenamefont {Yan}, \citenamefont {S\'anchez~S\'anchez},\ and\ \citenamefont
  {Valderrama}}]{Peng:2020hql}%
  \BibitemOpen
  \bibfield  {author} {\bibinfo {author} {\bibfnamefont {F.-Z.}\ \bibnamefont
  {Peng}}, \bibinfo {author} {\bibfnamefont {M.-J.}\ \bibnamefont {Yan}},
  \bibinfo {author} {\bibfnamefont {M.}~\bibnamefont {S\'anchez~S\'anchez}}, \
  and\ \bibinfo {author} {\bibfnamefont {M.~P.}\ \bibnamefont {Valderrama}},\
  }\href {\doibase 10.1140/epjc/s10052-021-09416-x} {\bibfield  {journal}
  {\bibinfo  {journal} {Eur. Phys. J. C}\ }\textbf {\bibinfo {volume} {81}},\
  \bibinfo {pages} {666} (\bibinfo {year} {2021})},\ \Eprint
  {http://arxiv.org/abs/2011.01915} {arXiv:2011.01915 [hep-ph]} \BibitemShut
  {NoStop}%
\bibitem [{\citenamefont {Chen}\ \emph {et~al.}(2021)\citenamefont {Chen},
  \citenamefont {Chen}, \citenamefont {Liu},\ and\ \citenamefont
  {Liu}}]{Chen:2020uif}%
  \BibitemOpen
  \bibfield  {author} {\bibinfo {author} {\bibfnamefont {H.-X.}\ \bibnamefont
  {Chen}}, \bibinfo {author} {\bibfnamefont {W.}~\bibnamefont {Chen}}, \bibinfo
  {author} {\bibfnamefont {X.}~\bibnamefont {Liu}}, \ and\ \bibinfo {author}
  {\bibfnamefont {X.-H.}\ \bibnamefont {Liu}},\ }\href {\doibase
  10.1140/epjc/s10052-021-09196-4} {\bibfield  {journal} {\bibinfo  {journal}
  {Eur. Phys. J. C}\ }\textbf {\bibinfo {volume} {81}},\ \bibinfo {pages} {409}
  (\bibinfo {year} {2021})},\ \Eprint {http://arxiv.org/abs/2011.01079}
  {arXiv:2011.01079 [hep-ph]} \BibitemShut {NoStop}%
\bibitem [{\citenamefont {Sakai}\ \emph {et~al.}(2019)\citenamefont {Sakai},
  \citenamefont {Jing},\ and\ \citenamefont {Guo}}]{Sakai:2019qph}%
  \BibitemOpen
  \bibfield  {author} {\bibinfo {author} {\bibfnamefont {S.}~\bibnamefont
  {Sakai}}, \bibinfo {author} {\bibfnamefont {H.-J.}\ \bibnamefont {Jing}}, \
  and\ \bibinfo {author} {\bibfnamefont {F.-K.}\ \bibnamefont {Guo}},\ }\href
  {\doibase 10.1103/PhysRevD.100.074007} {\bibfield  {journal} {\bibinfo
  {journal} {Phys. Rev. D}\ }\textbf {\bibinfo {volume} {100}},\ \bibinfo
  {pages} {074007} (\bibinfo {year} {2019})},\ \Eprint
  {http://arxiv.org/abs/1907.03414} {arXiv:1907.03414 [hep-ph]} \BibitemShut
  {NoStop}%
\bibitem [{\citenamefont {Du}\ \emph {et~al.}(2020{\natexlab{b}})\citenamefont
  {Du}, \citenamefont {Baru}, \citenamefont {Guo}, \citenamefont {Hanhart},
  \citenamefont {Mei\ss{}ner}, \citenamefont {Oller},\ and\ \citenamefont
  {Wang}}]{Du:2019pij}%
  \BibitemOpen
  \bibfield  {author} {\bibinfo {author} {\bibfnamefont {M.-L.}\ \bibnamefont
  {Du}}, \bibinfo {author} {\bibfnamefont {V.}~\bibnamefont {Baru}}, \bibinfo
  {author} {\bibfnamefont {F.-K.}\ \bibnamefont {Guo}}, \bibinfo {author}
  {\bibfnamefont {C.}~\bibnamefont {Hanhart}}, \bibinfo {author} {\bibfnamefont
  {U.-G.}\ \bibnamefont {Mei\ss{}ner}}, \bibinfo {author} {\bibfnamefont
  {J.~A.}\ \bibnamefont {Oller}}, \ and\ \bibinfo {author} {\bibfnamefont
  {Q.}~\bibnamefont {Wang}},\ }\href {\doibase 10.1103/PhysRevLett.124.072001}
  {\bibfield  {journal} {\bibinfo  {journal} {Phys. Rev. Lett.}\ }\textbf
  {\bibinfo {volume} {124}},\ \bibinfo {pages} {072001} (\bibinfo {year}
  {2020}{\natexlab{b}})},\ \Eprint {http://arxiv.org/abs/1910.11846}
  {arXiv:1910.11846 [hep-ph]} \BibitemShut {NoStop}%
\bibitem [{\citenamefont {Du}\ \emph {et~al.}(2021{\natexlab{b}})\citenamefont
  {Du}, \citenamefont {Baru}, \citenamefont {Guo}, \citenamefont {Hanhart},
  \citenamefont {Mei\ss{}ner}, \citenamefont {Oller},\ and\ \citenamefont
  {Wang}}]{Du:2021fmf}%
  \BibitemOpen
  \bibfield  {author} {\bibinfo {author} {\bibfnamefont {M.-L.}\ \bibnamefont
  {Du}}, \bibinfo {author} {\bibfnamefont {V.}~\bibnamefont {Baru}}, \bibinfo
  {author} {\bibfnamefont {F.-K.}\ \bibnamefont {Guo}}, \bibinfo {author}
  {\bibfnamefont {C.}~\bibnamefont {Hanhart}}, \bibinfo {author} {\bibfnamefont
  {U.-G.}\ \bibnamefont {Mei\ss{}ner}}, \bibinfo {author} {\bibfnamefont
  {J.~A.}\ \bibnamefont {Oller}}, \ and\ \bibinfo {author} {\bibfnamefont
  {Q.}~\bibnamefont {Wang}},\ }\href {\doibase 10.1007/JHEP08(2021)157}
  {\bibfield  {journal} {\bibinfo  {journal} {JHEP}\ }\textbf {\bibinfo
  {volume} {08}},\ \bibinfo {pages} {157} (\bibinfo {year}
  {2021}{\natexlab{b}})},\ \Eprint {http://arxiv.org/abs/2102.07159}
  {arXiv:2102.07159 [hep-ph]} \BibitemShut {NoStop}%
\bibitem [{\citenamefont {Aaij}\ \emph {et~al.}(2017)\citenamefont {Aaij} \emph
  {et~al.}}]{LHCb:2017iph}%
  \BibitemOpen
  \bibfield  {author} {\bibinfo {author} {\bibfnamefont {R.}~\bibnamefont
  {Aaij}} \emph {et~al.} (\bibinfo {collaboration} {LHCb}),\ }\href {\doibase
  10.1103/PhysRevLett.119.112001} {\bibfield  {journal} {\bibinfo  {journal}
  {Phys. Rev. Lett.}\ }\textbf {\bibinfo {volume} {119}},\ \bibinfo {pages}
  {112001} (\bibinfo {year} {2017})},\ \Eprint
  {http://arxiv.org/abs/1707.01621} {arXiv:1707.01621 [hep-ex]} \BibitemShut
  {NoStop}%
\bibitem [{\citenamefont {Workman}(2022)}]{Workman:2022ynf}%
  \BibitemOpen
  \bibfield  {author} {\bibinfo {author} {\bibfnamefont {R.~L.}\ \bibnamefont
  {Workman}} (\bibinfo {collaboration} {Particle Data Group}),\ }\href@noop {}
  {\bibfield  {journal} {\bibinfo  {journal} {PTEP}\ }\textbf {\bibinfo
  {volume} {2022}},\ \bibinfo {pages} {083C01} (\bibinfo {year}
  {2022})}\BibitemShut {NoStop}%
\bibitem [{\citenamefont {Chung}(1971)}]{Chung:1971ri}%
  \BibitemOpen
  \bibfield  {author} {\bibinfo {author} {\bibfnamefont {S.~U.}\ \bibnamefont
  {Chung}},\ }\href {\doibase 10.5170/CERN-1971-008} {\bibfield  {journal}
  {\bibinfo  {journal} {CERN-71-08}\ } (\bibinfo {year} {1971}),\
  10.5170/CERN-1971-008}\BibitemShut {NoStop}%
\end{thebibliography}%

\end{document}